\documentclass[reprint,superscriptaddress,amsmath,amssymb,aps,prb,longbibliography]{revtex4-2}
\usepackage[utf8]{inputenc}
\usepackage{graphicx}
\usepackage{dcolumn}% Align table columns on decimal point
\usepackage{bm}% bold math
\usepackage[colorlinks,citecolor=blue,linkcolor=blue,urlcolor=blue]{hyperref}
\usepackage[normalem]{ulem}

\begin{document}

\title{Dominant $p$-wave pairing induced by nearest-neighbor attraction in the square-lattice extended Hubbard model}

\author{Zhangkai Cao}
\thanks{These authors contributed equally.}
\affiliation{School of Science, Harbin Institute of Technology, Shenzhen, 518055, China}

\author{Jianyu Li}
\thanks{These authors contributed equally.}
\affiliation{School of Science, Harbin Institute of Technology, Shenzhen, 518055, China}
\affiliation{Shenzhen Key Laboratory of Advanced Functional Carbon Materials Research and Comprehensive Application, Shenzhen 518055, China.}
 
\author{Jiahao Su}
\affiliation{School of Science, Harbin Institute of Technology, Shenzhen, 518055, China}
\affiliation{Shenzhen Key Laboratory of Advanced Functional Carbon Materials Research and Comprehensive Application, Shenzhen 518055, China.}

\author{Tao Ying}
\email{taoying86@hit.edu.cn}
\affiliation{School of Physics, Harbin Institute of Technology, Harbin 150001, China}

\author{Ho-Kin Tang}
\email{denghaojian@hit.edu.cn}
\affiliation{School of Science, Harbin Institute of Technology, Shenzhen, 518055, China}
\affiliation{Shenzhen Key Laboratory of Advanced Functional Carbon Materials Research and Comprehensive Application, Shenzhen 518055, China.}

\date{\today}

\begin{abstract} 

The two-dimensional (2D) Hubbard model is widely believed to capture the key ingredients of high-temperature superconductivity in cuprate materials. Here, we report a constrained path quantum Monte Carlo study of the square-lattice extended Hubbard model with on-site Coulomb repulsion $U$ and nearest-neighbor (NN) electron attraction $V$. Upon doping $\delta$ = 0.125, we find that NN electron attraction $V$ significantly promotes an exotic spin-triplet ($p$-wave) pairing, with the $p$-wave pairing correlations growing stronger as $V$ increases. In the meanwhile, the $d_{x^2-y^2}$-wave ($d$-wave) pairing in the intermediate coupling regime shows insignificant response to the increase of $V$. The spin density wave order, which is observed near half-filling in the particle-hole channel, is also almost unaffected by the NN electron attraction $V$, reinforcing the established relationship between $d$-wave superconductivity (SC) and spin order. As the doping becomes heavier (i.e., $\delta$ increases), the dominant region of $p$-wave pairing expands, further suppressing the presence of $d$-wave pairing. Our work suggests the $p$-wave pairing region can be induced and further broadened by the NN electron attraction $V$, offering a feasible mechanism to realize $p$-wave SC in realistic cuprate materials.

\end{abstract} 
\maketitle

% \section{Introduction}

In almost four decades since the discovery of copper-based high-temperature ($T_c$) superconductors \cite{Bednorz1986}, later identified with a $d_{x^2-y^2}$-wave ($d$-wave) pairing symmetry \cite{Tsuei2000-fb}, reformed our understanding of strongly correlated electron systems, many of which defy Landau’s Fermi-liquid theory \cite{landau1957theory}. Lots of studies have pointed to the possibility of a phase transition \cite{Shekhter2013-vc} and symmetry breaking \cite{Xia2008-pa} especially in the underdoped region, resulting in many interweaving orders, including various forms of charge density wave (CDW) \cite{Comin2014-bz}, spin density wave (SDW) \cite{Moon2009-il}, pair density wave (PDW) \cite{Shi2020-vh}, electron nematic order \cite{Sato2017-bt}. The understanding of how these orders including superconducting, compete and cooperate with each other, has been the most significant problem in condensed matter physics for several decades remaining substantially unsolved \cite{Keimer2015-jp,Fradkin2015-xz}.

Although the ultimate theory for high-$T_c$ superconductivity remains controversial, the strong  electron repulsive interactions in Copper 3d orbitals are believed to play crucial roles \cite{Mai2021-wv}. The two-dimensional (2D) fermionic Hubbard model \cite{Zhang1988-dw,Arovas2022-ua,noauthor_2013-wp,qin2022hubbard} successfully captures most of the important physics, like antiferromagnetism at half filling \cite{Simons2015-jb} and competing orders in doped cuprates \cite{Scalapino2012-jr,Fradkin2015-xz,Zheng2017-of}. However, the search for a $d$-wave superconductivity (SC) phase in the pure 2D Hubbard Model has not been quite so successful \cite{qin2020absence}, which is still a lot of controversy. 
Recently, dome-like SC region is found in both the electron- and hole-doped regimes of the 2D Hubbard model with next-nearest-neighbor (NNN) hopping $t^{\prime}$ \cite{Xu2024-wt}.
Both positive and negative results have been observed for $d$-wave pairing order through various numerical simulation methods, reflecting the Hubbard model and its cousin $t-J$ model exhibit extreme sensitivity to ground state configurations and low-lying excitations, as well as the competition between $d$-wave SC and other orders \cite{Himeda2002-mc,Jiang2019-km,Xu2024-wt,Corboz2014-kr,Zheng2016-xt,qin2020absence}.

The relation between superconductivity and other orders is also strongly influenced by the variations of the Hubbard model, such as the inclusion of the nearest-neighbor (NN) attractive interaction $V$.
Moreover, recent angle-resolved photoemission spectroscopy (ARPES) experiments on one-dimensional (1D) cuprate chain Ba$_{2-x}$Sr$_x$CuO$_{3+\delta}$ \cite{Chen2021-xx} have revealed an anomalously sizable attractive interaction between NN electrons, possibly mediated by phonons \cite{Chen2021-xx,Wang2021-vs,Tang2023-aa}. Such an effective attraction, which has been largely overlooked in previous studies, could play a crucial role in understanding high-$T_c$ superconductivity. Although not so strong as the on-site Coulomb repulsion, this effective attractive interaction is comparable to the electron hopping integral ($V\sim-t$), and thereby also should not be ignored in 2D systems \cite{Wang2021-vs,Peng2023-on}.

The effect of the NN electron attraction $V$ in the 1D extended Hubbard chain has been examined recently \cite{Qu2022-yv}, which can host dominant spin-triplet ($p$-wave) pairing correlations and divergent superconductive susceptibility. This offers a feasible mechanism for realizing exotic $p$-wave SC in realistic cuprate chains. Beyond 1D, the density matrix renormalization group (DMRG) study of the extended Hubbard model on long four-leg cylinders on the square lattice finds that the NN electron attraction $V$ can notably enhance the long-distance superconducting correlations while simultaneously suppressing the CDW correlations \cite{Peng2023-on}.  Numerically exact diagonalization study of the Hamiltonian is currently restricted to 4 × 4 systems \cite{Chen2023-py}, so whether the prediction of a $p$-wave paired state could be stabilized in larger systems is doubtful. 

Moreover, the proposed competition between spin-singlet ($s$-wave and $d$-wave) pairing and spin-triplet ($p$-wave) pairing warrants further investigation \cite{Gukelberger2014-sf,Kallin2012-pl,Lee2008-kc,Guo2023-vr}. Although multiple possible triplet superconductors have been find, such as ${\rm Sr}_2{\rm RuO}_4$ \cite{Mackenzie2003-mn,Kinjo2022-vi}, ${\rm UPt}_3$ \cite{Sauls1994-vs}, ${\rm UTe}_2$ \cite{Jiao2020-ff}, ${\rm WTe}_2$ \cite{Jia2021-xw}, and ${\rm Cu_xBi_2Se_3}$ \cite{Matano_undated-ie}, the pairing symmetry and microscopic superconducting mechanism of these materials are not yet completely understood.  
In the majority of unconventional superconductors, such as copper oxide \cite{Comin2014-bz,Keimer2015-jp} and iron-based \cite{Liu2023-vc} high-$T_c$ superconductors, it is generally believed that the electron Cooper pairs exist in a spin-singlet state with a total spin $S=0$. Spin-triplet SC with $S=1$ possesses internal properties, leading to rich physics such as further possible spin-rotation symmetry breaking, collective modes of the order parameter and multiple phases of the condensate \cite{Mackenzie2003-mn,Kinjo2022-vi,Sauls1994-vs,Jiao2020-ff,Jia2021-xw,Matano_undated-ie}. Recently, the electrons with the same spin at the domain walls in the 1-UC Fe($\rm T_e, S_e$)/STO were found to form non-zero momentum Cooper pairs, thereby generating incommensurate PDW state \cite{Liu2023-vc}. This exotic triplet equal-spin pairing state can provide a new material platform to study the PDW state and its interplay with the topological electronic states and unconventional superconductivity.
% Furthermore, the notion of exploring the competition between trivial singlet pairing and non-trivial $p$-wave triplet pairing has emerged as a challenging and prominent area of investigation in recent research \cite{Gukelberger2014-sf,Kallin2012-pl,Lee2008-kc,Guo2023-vr,Crepel2022-be}. 
For $p$-wave superconductors, the disruption of time-reversal symmetry can give rise to topological superconductivity, which holds promising potential for applications in topological quantum computing \cite{Sarma2015-jq,Crepel2022-be}. 

Considering the structural and quantum chemistry similarities among cuprates, the NN attraction effect should not be ignored in 2D systems. Therefore, investigating the 2D extended Hubbard model with NN attraction could be crucial for understanding the high-$T_c$ pairing mechanism.  
Here, we using the constrained path quantum Monte Carlo~(CPQMC) \cite{Zhang1995-hn,Zhang1997-mr} to systematically study the phase diagrams of the square-lattice extended Hubbard model with on-site Coulomb repulsion $U$ and NN attraction $V$ of doped cuprates, so called $t-U-V$ model. CPQMC partially solves sign problem concerning only the ground state of the system at zero temperature. Firstly, we give a general phase diagram scan of the $U$-$V$ space at the doping $\delta$ = 0.125, where the $d$-wave and exotic $p$-wave triplet pairing phases exist. We find that $V$ can notably drive and enhance the exotic $p$-wave spin-triplet pairing phase, which also leads to the competition between $d$-wave and $p$-wave pairing. Besides the pairing phase, a SDW order exists near the half-filling in the particle-hole channel. Moreover, as doping increases, the dominant region of $p$-wave pairing expands, further suppressing the presence of $d$-wave pairing.
More technical details and additional results are included in the Supplemental Material \cite{Cao2024V-supp}.

\begin{figure}[b!]
    \centering 
    \includegraphics[width=1\linewidth]{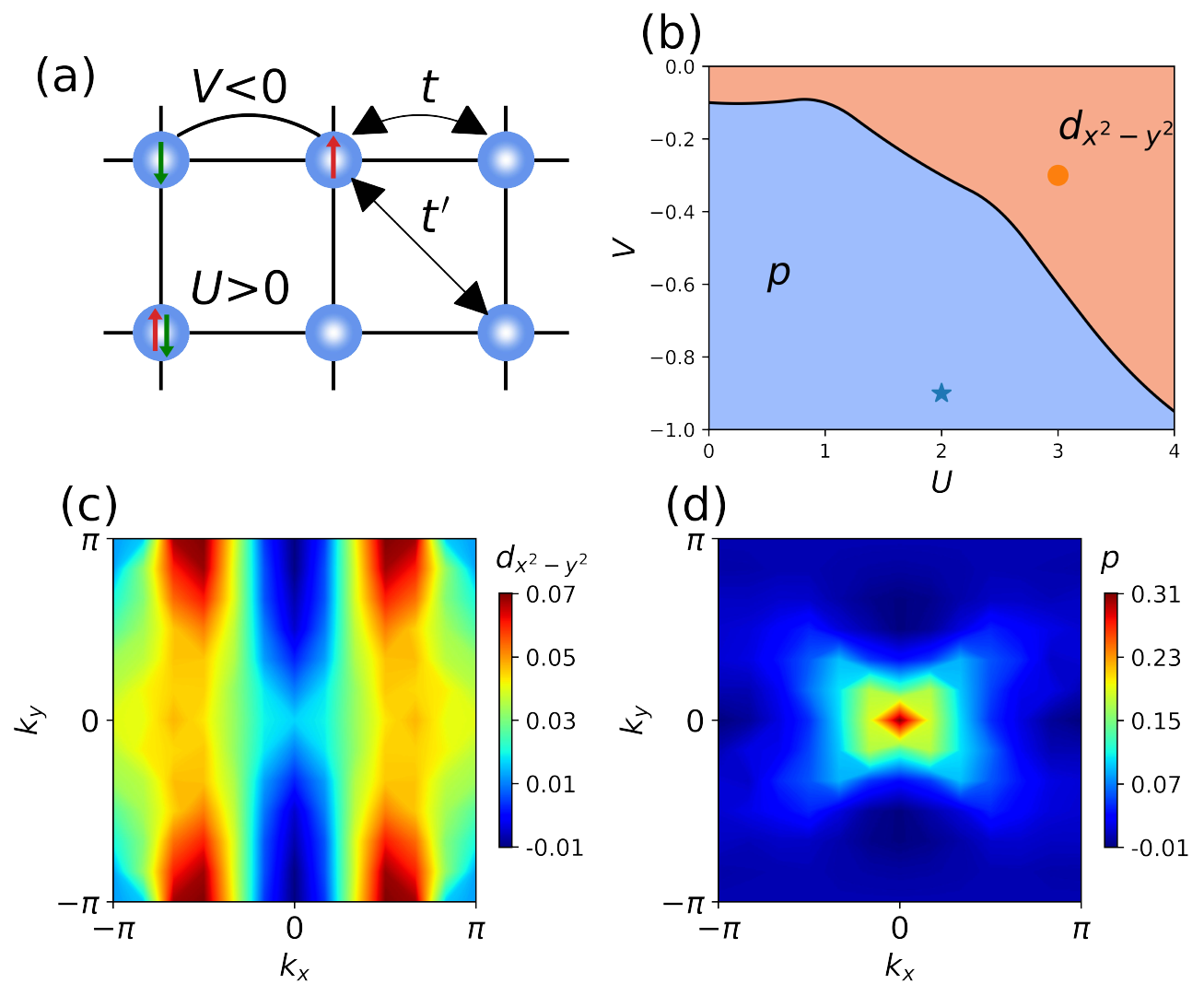}
    % \caption{(Color online) Model and phase diagram. (a) Illustration the $t-U-V$ Hubbard model with NN hopping $t$, NNN hopping $t^{\prime}$, on-site repulsive $U$, and NN attractive $V$ terms. (b) Schematic zero-temperature phase diagram at $\delta$ = 0.125, there are conventional $d$-wave and exotic $p$-wave triplet pairing phases. (c) The effective $d$-wave pair momentum distribution function $N^{\rm eff}_{\mathrm d_{x^2-y^2}-pair}({\bf k})$ at $\delta$= 0.125 with $U= 3$ and $V= -0.3$ on 12 $\times$ 12 lattice. (d) The effective $p$-wave pair momentum distribution function $N^{\rm eff}_{\mathrm p-pair}({\bf k})$ at $\delta$ = 0.125 with $U = 2$ and $V= -0.9$ on 12 $\times$ 12 lattice.   }
    \caption{(Color online) Model and phase diagram. (a) Illustration the $t-U-V$ Hubbard model with NN hopping $t$, NNN hopping $t^{\prime}$, on-site repulsive $U$, and NN attractive $V$ terms. (b) Schematic zero-temperature phase diagram at the underdoping case $\delta$ = 0.125, there are conventional $d$-wave and exotic $p$-wave triplet pairing phases. (c) The effective $d$-wave pair momentum distribution function $N^{\rm eff}_{d_{x^2-y^2}-pair}({\bf k})$ at $U= 3$ and $V= -0.3$, and (d) The effective $p$-wave pair momentum distribution function $N^{\rm eff}_{p-pair}({\bf k})$ at $U = 2$ and $V= -0.9$ on 12 $\times$ 12 lattice.}
    \label{fig1}
\end{figure}

We employ the CPQMC to investigate the ground-state properties of the hole-doped Hubbard model on the square lattice defined by the extended Hamiltonian
\begin{equation}\label{eq:hamiltonian}
\begin{aligned}
H =&-\sum_{\substack{i, j, \sigma}} \left( t_{i, j} c_{i,\sigma}^\dag c_{j,\sigma} + h.c. \right)
+ U \sum_i n_{i,\uparrow} n_{i,\downarrow} \\
&+ V \sum_{\substack{i, j, \sigma, \sigma'}} n_{i,\sigma} n_{j,\sigma'}
\end{aligned}
\end{equation}
where $c_{i,\sigma}^\dag$ ($c_{i,\sigma}^{\,}$) is electron creation (annihilation) operator with spin $\sigma$ = $\uparrow,\downarrow$, and $n_{i,\sigma}=c_{i,\sigma}^\dag c_{i,\sigma}$ is the electron number operator. We give an illustration of the $t-U-V$ Hubbard model on 2D square lattice (see Fig.\ \ref{fig1}(a)), where the electron hopping amplitude $t_{i, j} = t$ if $i$ and $j$ are the NN sites, and $t_{i, j} = t^{\prime}$ for NNN sites. Here, we assume $t=1$ for simplicity. We use $t^{\prime}=-0.2$ for hole doping, according to band-structure calculations in cuprates \cite{Andersen1995-pi,Hirayama2018-tm}. $U > 0$ is the on-site Coulomb repulsion and $V < 0$ is the NN electron attraction. We concentrate solely on the hole doped case, where the hole doping is given via $\delta$ = $1- N_e/N$ = $1-n$, where $N_e$ is the number of electrons in the system and $N$ is the number of lattice sites. 

We have defined the effective pair momentum distribution function
\begin{equation}
N^{\rm eff}_{\mathrm \zeta-pair}({\bf k}) = (1/N)\sum_{i,j} \mbox{exp}[i{\bf k}({\bf r}_i-{\bf r}_j)]C^{\rm eff}_{\mathrm \zeta-pair}(i,j),
\label{nspdtkpair}
\end{equation}
where $C^{\rm eff}_{\mathrm \zeta-pair}$ is the effective real-space correlation of the different pairing, and $\zeta =  s,\ d_{x^2-y^2},\ d_{xy},\ p$. Further detailed definitions of correlation functions can be found in the Supplemental Material \cite{Cao2024V-supp}.

% \section{results and discussion}

% \subsection{phase diagram}  

We focus on the electron density $n$ = 0.875 corresponding to the underdoping $\delta$ = 0.125, which is optimal for pairing to systematically explore its phase diagram. At zero temperature, we summarize our main findings in the phase diagram of the square-lattice extended Hubbard model in Fig.\ \ref{fig1}(b), where the conventional $d$-wave and exotic $p$-wave triplet pairing phases in the ground state is uncovered. At $\delta$ = 0.125, the $d_{x^2-y^2}$-wave pairing phase dominates the system when there is no attraction $V$.  As $V$ increases, a slight enhancement of NN electron attraction $V$ can drive the system to the $p$-wave triplet pairing phase, especially at relatively small $U$. But in the slightly larger $U$ region, a larger attraction $V$ is needed to induce the generation of $p$-wave pairing to suppress $d$-wave pairing.
Notably, in specific parameter regimes, fermions exhibit the potential to form pairs with exotic $p$-wave pairing symmetries, further contributing to this intricate pairing interplay \cite{Qu2022-yv}.This suggests the $p$-wave pairing order region is induced and further broadened by the NN electron attraction $V$.

Then we analyze the characteristic of the effective $d$-wave and $p$-wave pairing correlation function in momentum space. In Fig.\ \ref{fig1}(c), we show the effective $d$-wave pairing correlation function $N^{\rm eff}_{d_{x^2-y^2}-pair}({\bf k})$ at $\delta$ = 0.125 with $U= 3$ and $V= -0.3$, which exhibits a singular nonzero condensation structure, and its maximum value occurs near nonzero momentum point $Q=(2\pi/3,\ \pi)$. %This represents the uneven distribution of $d$-wave pairing in momentum space. %\zhang{PDW is an inhomogeneous superconducting state whose Cooper pairs poss a finite momentum \cite{Keimer2015-jp,Fradkin2015-xz,Shi2020-vh}.} 
This uneven distribution of $d$-wave pairing in momentum space is reminiscent of the PDW order, which is an inhomogeneous superconducting state whose Cooper pairs poss a finite momentum \cite{Keimer2015-jp,Fradkin2015-xz,Shi2020-vh}. In other words, we demonstrate the presence of the $d$-wave PDW phase.
Then, we focus on $p$-wave pairing state, which refers to the pairing between the same spin species. In Fig.\ \ref{fig1}(d), we show the effective $p$-wave pairing correlation function $N^{\rm eff}_{p-pair}({\bf k})$ at $\delta$= 0.125 with $U= 2$ and $V= -0.9$, which condenses at point $Q=(0,\ 0)$ and locates in the dominant region of weak-coupling $U$ and larger $V$, supporting the formation of Cooper-pair triplets with zero center-of-mass momentum and $p$-wave symmetry. We provide the effective $d$-wave and $p$-wave pairing correlation function in momentum space with the change of $U$, $V$ and $\delta$ in the Supplemental Material \cite{Cao2024V-supp}. %In the dominant region of the $d$-wave pairing, $d$-wave pairing correlation function exhibits a singular nonzero condensation structure, which demonstrate the presence of the $d$-wave PDW phase. However, in the dominant region of the $p$-wave pairing, d-wave pairing gradually evolves into $Q=(0,\ 0)$ condensation and has a lower strength than the $p$-wave pairing. The $p$-wave pairing strength increases with the increase of $V$ and suppressed by the increase of $U$. In the dominant region of the $p$-wave pairing, the p-wave pairing always condenses at point $Q=(0,\ 0)$. With doping increases, we found that the condensation point of the $d$-wave pairing becomes slightly dispersed and the intensity decreases, the $p$-wave pairing has always been $(0,\ 0)$ condensation and its intensity has remained basically unchanged.
In the region where $d$-wave pairing dominates, as discussed above, the $d$-wave pairing correlation function displays a singular nonzero condensation structure, indicating the presence of a $d$-wave PDW phase. In contrast, in the region where $p$-wave pairing dominates, the $d$-wave pairing gradually evolves into a $Q=(0,\ 0)$ condensation and has a lower strength than that of the $p$-wave pairing, with the latter consistently condenses at $Q=(0,\ 0)$. The $p$-wave pairing strength increases as $V$ is raised, while it is suppressed by the increase of $U$. As doping increases, we observe that the $d$-wave pairing condensation point becomes more dispersed, with its intensity decreasing, while the $p$-wave pairing continues to condense at $(0,\ 0)$, maintaining a relatively stable intensity.

\begin{figure}[!]
    \centering 
    \includegraphics[width=1\linewidth]{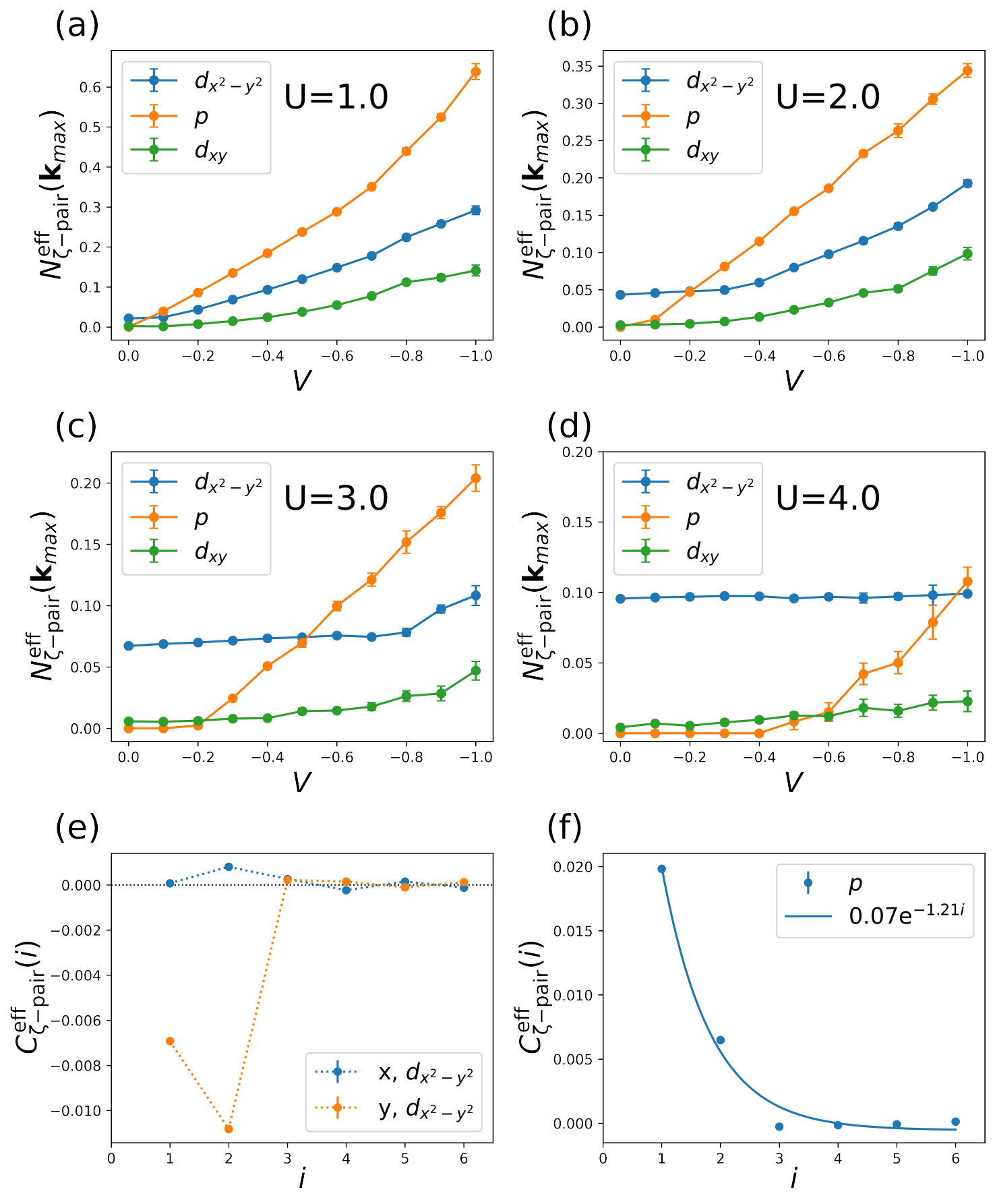}
    \caption{(Color online) The competition and distinction of $d$-wave and $p$-wave pairing at $\delta$ = 0.125 on 12 $\times$ 12 lattice. (a)-(d) The effective pair momentum distribution function $N^{\rm eff}_{d_{x^2-y^2}-pair}({\bf k})$, $N^{\rm eff}_{p-pair}({\bf k})$ and $N^{\rm eff}_{d_{xy}-pair}({\bf k})$ as a function of $V$, and fix (a) $U$ = 1.0, (b) $U$ = 2.0, (c) $U$ = 3.0, (d) $U$ = 4.0.  (e) The effective correlation function of $d$-wave pairing mode in real space in the $x$ and $y$ direction with $U = 3$ and $V = -0.3$, which displays a different staggered behavior with distance. (f) The effective correlation function of $p$-wave pairing mode in real space with $U = 2$ and $V = -0.9$, which displays an exponential decay with distance.  }
    \label{fig2}
\end{figure}

To show the competition and distinction of $d$-wave and $p$-wave pairing order, we present simulations along the vertical path in the phase diagram at $\delta$ = 0.125, namely, fixing $U$ = $1,\ 2,\ 3,\ 4$ as shown in Fig.\ \ref{fig2}(a)-(d). According to the results, the $d$-wave pairing exhibits different behaviors depending on the strength of $U$. We found that in the absence of $V$, the strength of the $d$-wave pairing in the system increases with the increase of $U$. As $V$ is enhanced, the strength of the $d$-wave pairing shows two types of changes: (i) in the weak coupling regime ($U = 1,\ 2$), the $d$-wave pairing slowly increases with the increase of $V$; (ii) in the intermediate coupling regime ($U = 3,\ 4$), the $d$-wave pairing remains relatively unchanged with the variation of $V$. %The intermediate $U$ regime indicates that the Coulomb interaction is large enough for correlated systems to prohibit two electrons from occupying the same site, and the CPQMC simulations could be very difficult with larger $U$. Therefore, the simulation results of intermediate $U$ provide a more accurate depiction of the behavior of cuprate superconductors. In the intermediate coupling regime, the $d$-wave \zhang{pairing} did not significantly increase with the increase of $V$, indicating that the $d$-wave \zhang{pairing} is not affected by $V$ in strongly correlated system. This is contrary to the recent numerical simulations \cite{Jiang2022-ok,Zhang2022-ej}, which suggest that $V$ can enhance the $d$-wave SC in intermediate or even strong $U$.
% The second conclusion contrasts with recent numerical simulations \cite{Jiang2022-ok, Zhang2022-ej}, which suggest that $V$ can enhance $d$-wave pairing for intermediate or even stronger $U$. 
Although investigating stronger $U$ may provide further insight, such simulations are challenging for the CPQMC method. Consequently, our analysis primarily focuses on the intermediate $U$ regime, where the Coulomb interaction is already large enough for correlated systems to prohibit two electrons from occupying the same site, and is believed to capture the behavior of cuprate superconductors.

The $p$-wave pairing is entirely absent when $V=0$, indicating that the $p$-wave pairing is solely induced by the NN electron attraction $V$. In the weakly coupling regime, the $p$-wave pairing strength increases sharply with the increase of $V$ and quickly surpasses the strength of the $d$-wave pairing. In the intermediate coupling regime, the $p$-wave pairing is suppressed until $V$ reaches a certain strength after which it rapidly increases. Overall, the region dominated by $p$-wave pairing gradually expands and the strength of $p$-wave pairing gets significantly enhanced as $V$ increases. The competition between $d$-wave and $p$-wave pairing states at different $V$ is primarily driven by the variation of the strength of the $p$-wave pairing. We also investigated the effective $d_{xy}$-wave pairing correlation function and found that the strength of $d_{xy}$-wave remains relatively weak at $\delta$ = 0.125, suggesting that the main competition is between the $d$-wave and $p$-wave pairing in the system.

Then we analyze the characteristic of $d$-wave and $p$-wave pairing correlation function in real space. In Fig.\ \ref{fig2}(e), we show the effective $d$-wave correlation function in real space, which displays different staggered behaviors in the $x$ and $y$ direction with distance. Such different staggered behaviors represent the uneven distribution of $d$-wave pairing in real space, which is consistent with the $d$-wave pairing condensation at nonzero momentum point $Q=(2\pi/3,\ \pi)$ in momentum space, indicating that the $d$-wave pairing order spontaneously breaks the $C_4$ rotational symmetry of the system. This may be related to the origin of electron nematic order, which corresponds to that the electronic structure  preserves the translation symmetry but breaks the rotation symmetry of the CuO$_2$ plane. On the contrary, the real space $p$-wave pairing correlation function exhibits an exponential decay as a function of the distance and drops to 0 quickly, indicating that the $p$-wave pairing correlation is short-range, as shown in Fig.\ \ref{fig2}(f).

\begin{figure}[t!]
    \centering 
    \includegraphics[width=1\linewidth]{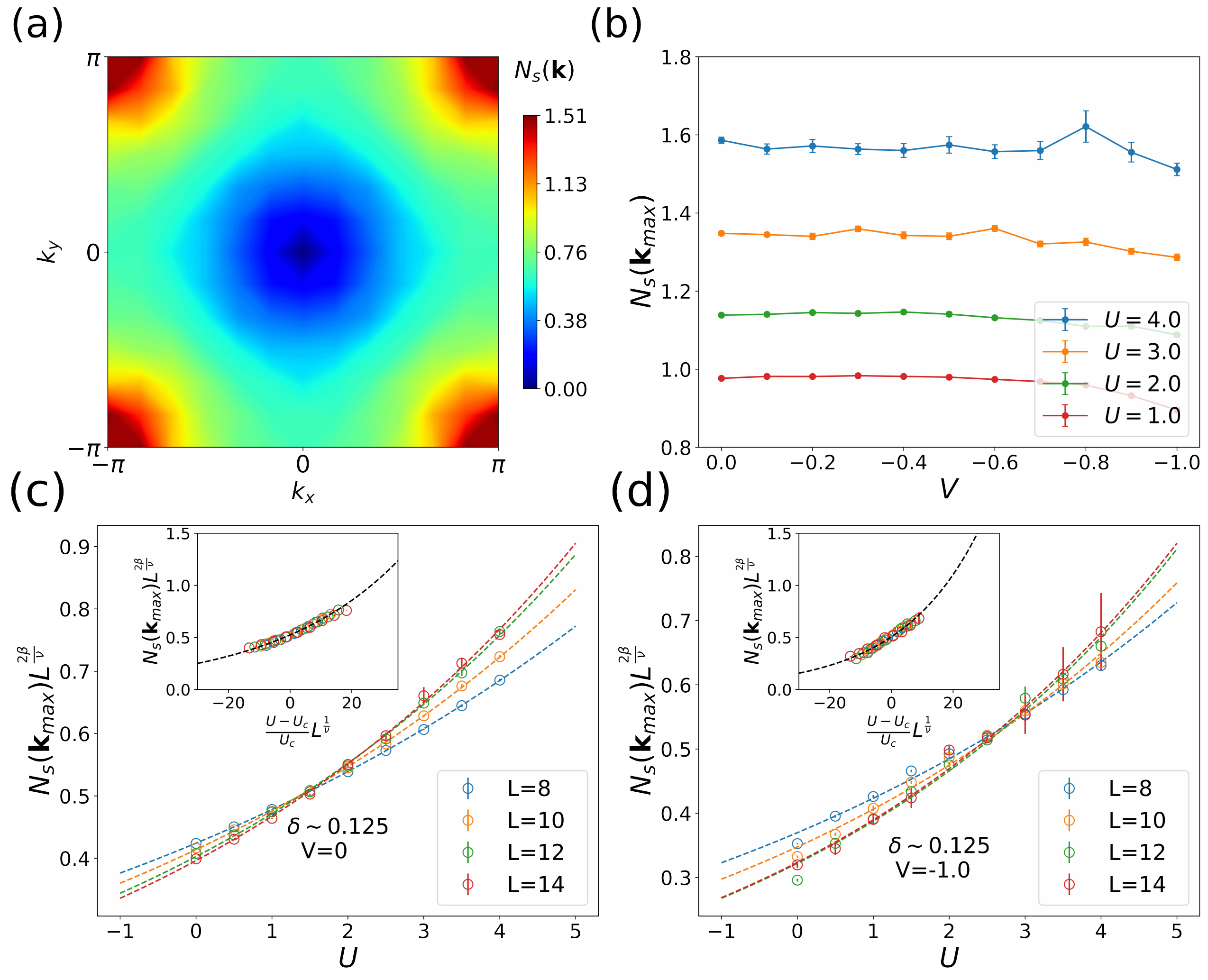}
    \caption{(Color online) Spin density correlation. (a) The spin structure factor $N_{S}({\bf k})$ at $\delta$ = 0.125 with $U = 4$ and $V= -1.0$ on 12 $\times$ 12 lattice. $N_{S}({\bf k})$ is peaked at the antiferromagnetic ordering momentum, ${\bf k}_{\rm max}$ = ($\pi$, $\pi$). (b) $N_{S}({\bf k_{max}})$ at $\delta$ = 0.125 as a function of $V$, with $U$ = 1.0, 2.0, 3.0, 4.0. Scaled spin structure factor $N_{S}({\bf k}_{\rm max})L^{\frac{2\beta}{\nu}}/N$ for different lattice sizes $L$ = 8, 10, 12 and 14, with (c) $V=0$ and (d) $V=-1.0$ at $\delta$ $\sim$ 0.125. The crossing of the different data sets at $U_c$ $\simeq$ 1.7 and $U_c$ $\simeq$ 2.6 respectively,  approximately gives the critical interaction $U_c$ identifying the phase transition. (Inset) Additionally, when scaling the onsite interaction $U$ by $(U- U_c)L^{\frac{1}{\nu}}/U_c$, the data sets for different system sizes collapse for critical exponents $\nu$ = 1.02(2) and $\beta$ = 0.85(2). }
    \label{fig3}
\end{figure}

We also studied the spin density correlation in the particle-hole channel, which reflects the magnetic property of the system. We defined the spin structure factor as $N_{S}({\bf k}) = (1/N)\sum_{i,j} \mbox{exp}[i{\bf k}({\bf r}_i-{\bf r}_j)]\, \langle {\bf S}_i \cdot{\bf S}_j \rangle,$ and ${\bf S}_i$ the spin operator at site $i$. At half filling or near half filling, the attractive Hubbard model is dominated by the ($\pi$, $\pi$) CDW. In contrast, a predominant positive $U > 0$ is expected to support the ($\pi$, $\pi$) SDW. 
We analyze the characteristic of spin structure factor $N_{S}({\bf k})$ in momentum space at $\delta$ = 0.125 with $U = 4$ and $V= -1.0$, as shown in Fig.\ \ref{fig3}(a). The SDW order, characterized by the momentum ($\pi$, $\pi$), leds to a long-range order of staggered spin density in real space.
To clearly show the impact of $V$ on SDW, we present the simulations of $N_{S}({\bf k})$ at $\delta$ = 0.125 as a function of $V$, namely, fixing $U$ = $1,\ 2,\ 3,\ 4$ as shown in Fig.\ \ref{fig3}(b).
The strength of spin correlation increases with the increase of $U$, which is consistent with the variation of $d$-wave pairing with $U$. As $V$ is enhanced, the strength of spin correlation remains basically unchanged at small $V$, but slightly decreases at large $V$. Such behavior indicates that the spin correlation is not significantly affected by the change of $V$, which is again similar to the behavior of the $d$-wave pairing in the intermediate coupling regime, to some extent proves the validness of the antiferromagnetic spin fluctuations as the pairing glue of the $d$-wave superconductivity \cite{Dahm2009-wg,Dong2022-xh}.

% In Fig.\ \ref{fig3}(a), we shows the CPQMC results together with a finite-size extrapolation at $\delta$ $\sim$ 0.125 and inset is $\delta$ $\sim$ 0.014. The SDW order appears at $\delta$ $\sim$ 0.014, the extrapolated value is greater than 0. On the contrary, when $\delta$ $\sim$ 0.125, SDW does not exist and the extrapolated value is less than 0. In the intermediate coupling regime ($U = 4$), doping disrupts the SDW, resulting in the absence of SDW in the system when away from the half filling. When SDW order is absent, $d$-wave \zhang{pairing} began to emerge and dominated the system. So on the other hand, it can also be said that $d$-wave \zhang{pairing order} disrupts the formation of SDW order.

As demonstrated in Fig.\ \ref{fig3}(c) and (d), we can approximately determined $U_c$ from the crossing point of the scaled spin structure factor $N_{S}({\bf k}_{\rm max})L^{\frac{2\beta}{\nu}}/N$ obtained for different lattice sizes L at $\delta$ $\sim$ 0.125 with $V = 0$ and $V = -1.0$, separately. The critical value $U = U_c$ represents a quantum phase transition of the system from a metal phase to the SDW phase. The critical exponents $\nu$ and $\beta$ in the scaling law are determined by collapsing the scaled $N_{S}({\bf k})$ arising from different sizes (inset of Fig.\ \ref{fig3}(c) and (d)), consistent with previous research in square-lattice \cite{Otsuka2020-wc}. 
% This phase transition was predicted to be of the Gross-Neveu universality class.
We found that at $V = 0$ and $V = -1.0$, the critical interactions are $U_c \simeq 1.7$ and $U_c \simeq$ 2.4 respectively, indicating that, in the presence of an attractive $V$, the system requires a larger critical $U$ to cause the phase transition.

\begin{figure}[!]
    \centering 
    \includegraphics[width=1\linewidth]{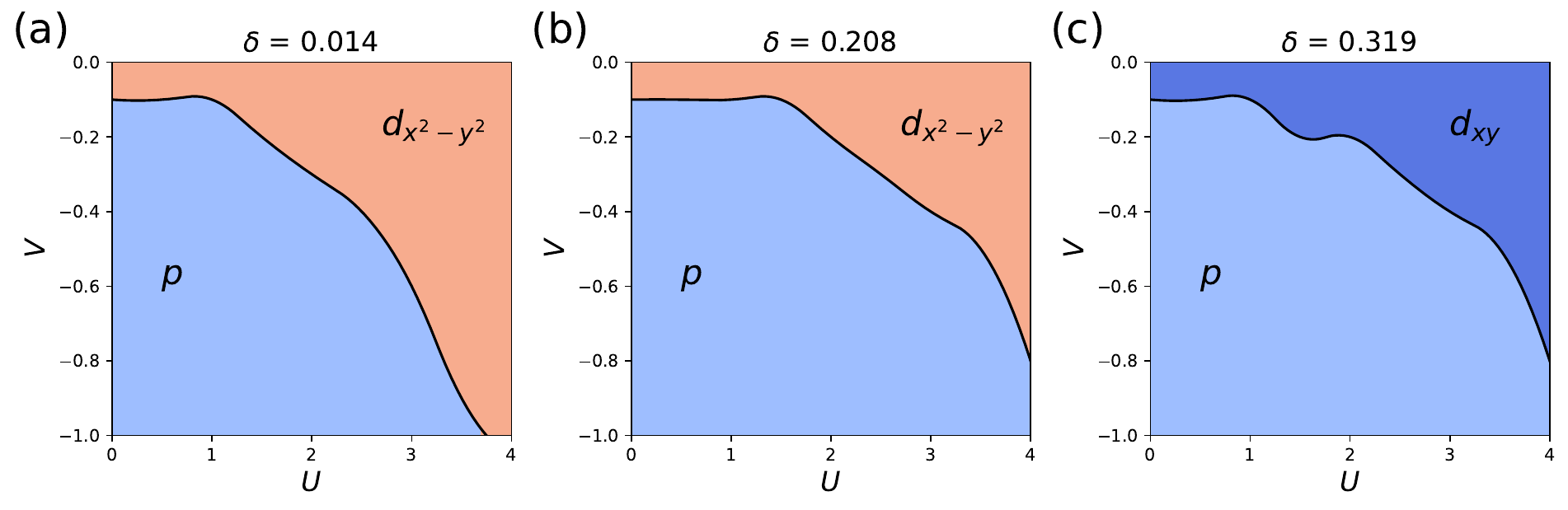}
    \caption{(Color online) Schematic zero-temperature phase diagram of different pairing modes under different doping: (a) $\delta$ = 0.014, (b) $\delta$ = 0.208, (c) $\delta$ = 0.319 on 12 $\times$ 12 lattice, $d_{x^2-y^2}$-wave, $d_{xy}$-wave and exotic $p$-wave triplet pairing phases are observed. }
    \label{fig4}
\end{figure}

Lastly, we explore the effect of doping on the pairing order. Three additional doping concentrations are considered, namely a lightly doping $\delta$ = 0.014 (near half filling), an overdoped case $\delta$ = 0.208, and a heavily overdoped case $\delta$ = 0.319. The zero-temperature phase diagrams are displayed in Fig.\ \ref{fig4}, with $d$-wave, $d_{xy}$-wave and exotic $p$-wave triplet pairing phases observed. At near half filling ($\delta$ = 0.014, Fig.\ \ref{fig4}(a)), underdoped ($\delta$ = 0.125, Fig.\ \ref{fig1}b) and slightly overdoped ($\delta$ = 0.208, Fig.\ \ref{fig4}(b)), the $d$-wave pairing phase dominates when there is no attraction $V$. However, in heavily overdoped case ($\delta$ = 0.319, Fig.\ \ref{fig4}(c)), the system becomes dominated by $d_{xy}$-wave pairing phase. This is consistent with Fig.~S4, where we give the zero-temperature phase diagram of the dominant pairing channel in $U-n$ space with $V$=0~\cite{Cao2024V-supp}.

Notably, there is no $p$-wave pairing phase in the absence of $V$ for all the doping cases. However, when $V$ presents, the $p$-wave triplet pairing order begins to competes with the $d$-wave pairing order (or the $d_{xy}$-wave pairing for the heavily doping case $\delta$ = 0.319), and overwhelms with an appropriate increase of $V$. Moreover, as the doping increases, the dominant region of $p$-wave pairing expands, further suppressing the presence of $d$-wave (or the $d_{xy}$-wave) pairing. Our results indicate that the $p$-wave pairing order is induced by the NN electron attraction $V$ and further benefits when the doping is heavier.
From the known results obtained in 1D cuprate chains, the critical strengths of $V$ for entering the $p$-wave pairing phase can be determined analytically at quarter filling in the $U \rightarrow \infty$ limit \cite{Luther1975-jd,Schulz1990-tb,Qu2022-yv}, in which the Luttinger parameter $K_{\rho}=1/[2 +(4/\pi){\rm sin}^{-1}(V/2)]$ exceeding the unity indicates that it has entered the $p$-wave pairing phase. In a 2D system, similar mechanism is probable to happen that the strong NN electron attraction $V$ drives the system towards the $p$-wave pairing phase.

% \section{SUMMARY}

In conclusion, we adopted CPQMC calculations of the 2D extended Hubbard model with the NN attraction $V$ to study the impact of the additional attractive $V$ on the pairing phase and density wave state. The quantum phase diagram containing $d$-wave, $d_{xy}$-wave and $p$-wave pairing phases at different doping is explored. We find that the $p$-wave pairing phase is completely induced by $V$, and the $p$-wave pairing strength is enhanced as $V$ increases. However, in the intermediate coupling regime, the $d$-wave pairing shows insignificant response to the increase of $V$. 
The spin density wave order, which is observed near half-filling in the particle-hole channel, is also almost unaffected by the NN electron attraction $V$, showing a similar behavior to the $d$-wave pairing with $V$. In a way, this supports that the spin fluctuations act as the dominant microscopic mechanism to drive the unconventional $d$-wave pairing phase. 
% Therefore, the NN attraction $V$, which is believed to be the main driving force of the $p$-wave pairing order from this study, may be argued to arise from the electron-phonon couplings \cite{Chen2021-xx,Wang2021-vs,Tang2023-aa}, independent of the spin correlation.}
As $\delta$ increases, the dominant region of $p$-wave pairing expands, further suppressing the presence of $d$-wave pairing.
Our work suggests the $p$-wave pairing region can be induced and further broadened by the NN electron attraction $V$, which will not only deepen the understanding of the high-$T_c$ pairing microscopical mechanism, but also offering a feasible mechanism to realize $p$-wave triplet SC \cite{Das2020-yp,Lu2014-bn} in realistic doped cuprates.

We acknowledge useful discussion with Ji Liu.
This work is supported by the National Natural Science Foundation of China~(Grant No. 12204130), Shenzhen Start-Up Research Funds~(Grant No. HA11409065), HITSZ Start-Up Funds~(Grant No. X2022000), Shenzhen Key Laboratory of Advanced Functional CarbonMaterials Research and Comprehensive Application (Grant No. ZDSYS20220527171407017). T.Y. acknowledges supports from Natural Science Foundation of Heilongjiang Province~(No.~YQ2023A004).

\bibliography{ref}

\begin{thebibliography}{59}%
\makeatletter
\providecommand \@ifxundefined [1]{%
 \@ifx{#1\undefined}
}%
\providecommand \@ifnum [1]{%
 \ifnum #1\expandafter \@firstoftwo
 \else \expandafter \@secondoftwo
 \fi
}%
\providecommand \@ifx [1]{%
 \ifx #1\expandafter \@firstoftwo
 \else \expandafter \@secondoftwo
 \fi
}%
\providecommand \natexlab [1]{#1}%
\providecommand \enquote  [1]{``#1''}%
\providecommand \bibnamefont  [1]{#1}%
\providecommand \bibfnamefont [1]{#1}%
\providecommand \citenamefont [1]{#1}%
\providecommand \href@noop [0]{\@secondoftwo}%
\providecommand \href [0]{\begingroup \@sanitize@url \@href}%
\providecommand \@href[1]{\@@startlink{#1}\@@href}%
\providecommand \@@href[1]{\endgroup#1\@@endlink}%
\providecommand \@sanitize@url [0]{\catcode `\\12\catcode `\$12\catcode `\&12\catcode `\#12\catcode `\^12\catcode `\_12\catcode `\%12\relax}%
\providecommand \@@startlink[1]{}%
\providecommand \@@endlink[0]{}%
\providecommand \url  [0]{\begingroup\@sanitize@url \@url }%
\providecommand \@url [1]{\endgroup\@href {#1}{\urlprefix }}%
\providecommand \urlprefix  [0]{URL }%
\providecommand \Eprint [0]{\href }%
\providecommand \doibase [0]{https://doi.org/}%
\providecommand \selectlanguage [0]{\@gobble}%
\providecommand \bibinfo  [0]{\@secondoftwo}%
\providecommand \bibfield  [0]{\@secondoftwo}%
\providecommand \translation [1]{[#1]}%
\providecommand \BibitemOpen [0]{}%
\providecommand \bibitemStop [0]{}%
\providecommand \bibitemNoStop [0]{.\EOS\space}%
\providecommand \EOS [0]{\spacefactor3000\relax}%
\providecommand \BibitemShut  [1]{\csname bibitem#1\endcsname}%
\let\auto@bib@innerbib\@empty
%</preamble>
\bibitem [{\citenamefont {Bednorz}\ and\ \citenamefont {M{\"u}ller}(1986)}]{Bednorz1986}%
  \BibitemOpen
  \bibfield  {author} {\bibinfo {author} {\bibfnamefont {J.~G.}\ \bibnamefont {Bednorz}}\ and\ \bibinfo {author} {\bibfnamefont {K.~A.}\ \bibnamefont {M{\"u}ller}},\ }\bibfield  {title} {\bibinfo {title} {Possible high ${T}_{c}$ superconductivity in the {Ba}$-${La}$-${Cu}$-${O} system},\ }\href {https://doi.org/10.1007/BF01303701} {\bibfield  {journal} {\bibinfo  {journal} {Z Physik B}\ }\textbf {\bibinfo {volume} {64}},\ \bibinfo {pages} {189} (\bibinfo {year} {1986})}\BibitemShut {NoStop}%
\bibitem [{\citenamefont {Tsuei}\ and\ \citenamefont {Kirtley}(2000)}]{Tsuei2000-fb}%
  \BibitemOpen
  \bibfield  {author} {\bibinfo {author} {\bibfnamefont {C.~C.}\ \bibnamefont {Tsuei}}\ and\ \bibinfo {author} {\bibfnamefont {J.~R.}\ \bibnamefont {Kirtley}},\ }\bibfield  {title} {\bibinfo {title} {{Pairing symmetry in cuprate superconductors}},\ }\href {https://doi.org/10.1103/RevModPhys.72.969} {\bibfield  {journal} {\bibinfo  {journal} {Reviews of modern physics}\ }\textbf {\bibinfo {volume} {72}},\ \bibinfo {pages} {969} (\bibinfo {year} {2000})}\BibitemShut {NoStop}%
\bibitem [{\citenamefont {Landau}(1957)}]{landau1957theory}%
  \BibitemOpen
  \bibfield  {author} {\bibinfo {author} {\bibfnamefont {L.~D.}\ \bibnamefont {Landau}},\ }\bibfield  {title} {\bibinfo {title} {The theory of a fermi liquid},\ }\href@noop {} {\bibfield  {journal} {\bibinfo  {journal} {Sov. Phys. JETP}\ }\textbf {\bibinfo {volume} {3}},\ \bibinfo {pages} {920} (\bibinfo {year} {1957})}\BibitemShut {NoStop}%
\bibitem [{\citenamefont {Shekhter}\ \emph {et~al.}(2013)\citenamefont {Shekhter}, \citenamefont {Ramshaw}, \citenamefont {Liang}, \citenamefont {Hardy}, \citenamefont {Bonn}, \citenamefont {Balakirev}, \citenamefont {McDonald}, \citenamefont {Betts}, \citenamefont {Riggs},\ and\ \citenamefont {Migliori}}]{Shekhter2013-vc}%
  \BibitemOpen
  \bibfield  {author} {\bibinfo {author} {\bibfnamefont {A.}~\bibnamefont {Shekhter}}, \bibinfo {author} {\bibfnamefont {B.~J.}\ \bibnamefont {Ramshaw}}, \bibinfo {author} {\bibfnamefont {R.}~\bibnamefont {Liang}}, \bibinfo {author} {\bibfnamefont {W.~N.}\ \bibnamefont {Hardy}}, \bibinfo {author} {\bibfnamefont {D.~A.}\ \bibnamefont {Bonn}}, \bibinfo {author} {\bibfnamefont {F.~F.}\ \bibnamefont {Balakirev}}, \bibinfo {author} {\bibfnamefont {R.~D.}\ \bibnamefont {McDonald}}, \bibinfo {author} {\bibfnamefont {J.~B.}\ \bibnamefont {Betts}}, \bibinfo {author} {\bibfnamefont {S.~C.}\ \bibnamefont {Riggs}},\ and\ \bibinfo {author} {\bibfnamefont {A.}~\bibnamefont {Migliori}},\ }\bibfield  {title} {\bibinfo {title} {{Bounding the pseudogap with a line of phase transitions in YBa$_{2}$Cu$_{3}$O$_{6+\delta}$}},\ }\href {https://doi.org/10.1038/nature12165} {\bibfield  {journal} {\bibinfo  {journal} {Nature}\ }\textbf {\bibinfo {volume} {498}},\ \bibinfo {pages} {75} (\bibinfo {year} {2013})}\BibitemShut {NoStop}%
\bibitem [{\citenamefont {Xia}\ \emph {et~al.}(2008)\citenamefont {Xia}, \citenamefont {Schemm}, \citenamefont {Deutscher}, \citenamefont {Kivelson}, \citenamefont {Bonn}, \citenamefont {Hardy}, \citenamefont {Liang}, \citenamefont {Siemons}, \citenamefont {Koster}, \citenamefont {Fejer},\ and\ \citenamefont {Kapitulnik}}]{Xia2008-pa}%
  \BibitemOpen
  \bibfield  {author} {\bibinfo {author} {\bibfnamefont {J.}~\bibnamefont {Xia}}, \bibinfo {author} {\bibfnamefont {E.}~\bibnamefont {Schemm}}, \bibinfo {author} {\bibfnamefont {G.}~\bibnamefont {Deutscher}}, \bibinfo {author} {\bibfnamefont {S.~A.}\ \bibnamefont {Kivelson}}, \bibinfo {author} {\bibfnamefont {D.~A.}\ \bibnamefont {Bonn}}, \bibinfo {author} {\bibfnamefont {W.~N.}\ \bibnamefont {Hardy}}, \bibinfo {author} {\bibfnamefont {R.}~\bibnamefont {Liang}}, \bibinfo {author} {\bibfnamefont {W.}~\bibnamefont {Siemons}}, \bibinfo {author} {\bibfnamefont {G.}~\bibnamefont {Koster}}, \bibinfo {author} {\bibfnamefont {M.~M.}\ \bibnamefont {Fejer}},\ and\ \bibinfo {author} {\bibfnamefont {A.}~\bibnamefont {Kapitulnik}},\ }\bibfield  {title} {\bibinfo {title} {{Polar Kerr-effect measurements of the high-temperature YBa$_{2}$Cu$_{3}$O$_{6+x}$ superconductor: evidence for broken symmetry near the pseudogap temperature}},\ }\href {https://doi.org/10.1103/PhysRevLett.100.127002} {\bibfield  {journal} {\bibinfo
  {journal} {Physical review letters}\ }\textbf {\bibinfo {volume} {100}},\ \bibinfo {pages} {127002} (\bibinfo {year} {2008})}\BibitemShut {NoStop}%
\bibitem [{\citenamefont {Comin}\ \emph {et~al.}(2014)\citenamefont {Comin}, \citenamefont {Frano}, \citenamefont {Yee}, \citenamefont {Yoshida}, \citenamefont {Eisaki}, \citenamefont {Schierle}, \citenamefont {Weschke}, \citenamefont {Sutarto}, \citenamefont {He}, \citenamefont {Soumyanarayanan}, \citenamefont {He}, \citenamefont {Le~Tacon}, \citenamefont {Elfimov}, \citenamefont {Hoffman}, \citenamefont {Sawatzky}, \citenamefont {Keimer},\ and\ \citenamefont {Damascelli}}]{Comin2014-bz}%
  \BibitemOpen
  \bibfield  {author} {\bibinfo {author} {\bibfnamefont {R.}~\bibnamefont {Comin}}, \bibinfo {author} {\bibfnamefont {A.}~\bibnamefont {Frano}}, \bibinfo {author} {\bibfnamefont {M.~M.}\ \bibnamefont {Yee}}, \bibinfo {author} {\bibfnamefont {Y.}~\bibnamefont {Yoshida}}, \bibinfo {author} {\bibfnamefont {H.}~\bibnamefont {Eisaki}}, \bibinfo {author} {\bibfnamefont {E.}~\bibnamefont {Schierle}}, \bibinfo {author} {\bibfnamefont {E.}~\bibnamefont {Weschke}}, \bibinfo {author} {\bibfnamefont {R.}~\bibnamefont {Sutarto}}, \bibinfo {author} {\bibfnamefont {F.}~\bibnamefont {He}}, \bibinfo {author} {\bibfnamefont {A.}~\bibnamefont {Soumyanarayanan}}, \bibinfo {author} {\bibfnamefont {Y.}~\bibnamefont {He}}, \bibinfo {author} {\bibfnamefont {M.}~\bibnamefont {Le~Tacon}}, \bibinfo {author} {\bibfnamefont {I.~S.}\ \bibnamefont {Elfimov}}, \bibinfo {author} {\bibfnamefont {J.~E.}\ \bibnamefont {Hoffman}}, \bibinfo {author} {\bibfnamefont {G.~A.}\ \bibnamefont {Sawatzky}}, \bibinfo {author} {\bibfnamefont
  {B.}~\bibnamefont {Keimer}},\ and\ \bibinfo {author} {\bibfnamefont {A.}~\bibnamefont {Damascelli}},\ }\bibfield  {title} {\bibinfo {title} {{Charge order driven by Fermi-arc instability in Bi$_{3}$Sr$_{2-x}$La$_{x}$CuO$_{6+\delta}$}},\ }\href {https://doi.org/10.1126/science.1242996} {\bibfield  {journal} {\bibinfo  {journal} {Science}\ }\textbf {\bibinfo {volume} {343}},\ \bibinfo {pages} {390} (\bibinfo {year} {2014})}\BibitemShut {NoStop}%
\bibitem [{\citenamefont {Moon}\ and\ \citenamefont {Sachdev}(2009)}]{Moon2009-il}%
  \BibitemOpen
  \bibfield  {author} {\bibinfo {author} {\bibfnamefont {E.~G.}\ \bibnamefont {Moon}}\ and\ \bibinfo {author} {\bibfnamefont {S.}~\bibnamefont {Sachdev}},\ }\bibfield  {title} {\bibinfo {title} {{Competition between spin density wave order and superconductivity in the underdoped cuprates}},\ }\href {https://doi.org/10.1103/PhysRevB.80.035117} {\bibfield  {journal} {\bibinfo  {journal} {Physical review. B, Condensed matter}\ }\textbf {\bibinfo {volume} {80}},\ \bibinfo {pages} {035117} (\bibinfo {year} {2009})}\BibitemShut {NoStop}%
\bibitem [{\citenamefont {Shi}\ \emph {et~al.}(2020)\citenamefont {Shi}, \citenamefont {Baity}, \citenamefont {Terzic}, \citenamefont {Sasagawa},\ and\ \citenamefont {Popović}}]{Shi2020-vh}%
  \BibitemOpen
  \bibfield  {author} {\bibinfo {author} {\bibfnamefont {Z.}~\bibnamefont {Shi}}, \bibinfo {author} {\bibfnamefont {P.~G.}\ \bibnamefont {Baity}}, \bibinfo {author} {\bibfnamefont {J.}~\bibnamefont {Terzic}}, \bibinfo {author} {\bibfnamefont {T.}~\bibnamefont {Sasagawa}},\ and\ \bibinfo {author} {\bibfnamefont {D.}~\bibnamefont {Popović}},\ }\bibfield  {title} {\bibinfo {title} {{Pair density wave at high magnetic fields in cuprates with charge and spin orders}},\ }\href {https://doi.org/10.1038/s41467-020-17138-z} {\bibfield  {journal} {\bibinfo  {journal} {Nature communications}\ }\textbf {\bibinfo {volume} {11}},\ \bibinfo {pages} {3323} (\bibinfo {year} {2020})}\BibitemShut {NoStop}%
\bibitem [{\citenamefont {Sato}\ \emph {et~al.}(2017)\citenamefont {Sato}, \citenamefont {Kasahara}, \citenamefont {Murayama}, \citenamefont {Kasahara}, \citenamefont {Moon}, \citenamefont {Nishizaki}, \citenamefont {Loew}, \citenamefont {Porras}, \citenamefont {Keimer}, \citenamefont {Shibauchi},\ and\ \citenamefont {Matsuda}}]{Sato2017-bt}%
  \BibitemOpen
  \bibfield  {author} {\bibinfo {author} {\bibfnamefont {Y.}~\bibnamefont {Sato}}, \bibinfo {author} {\bibfnamefont {S.}~\bibnamefont {Kasahara}}, \bibinfo {author} {\bibfnamefont {H.}~\bibnamefont {Murayama}}, \bibinfo {author} {\bibfnamefont {Y.}~\bibnamefont {Kasahara}}, \bibinfo {author} {\bibfnamefont {E.-G.}\ \bibnamefont {Moon}}, \bibinfo {author} {\bibfnamefont {T.}~\bibnamefont {Nishizaki}}, \bibinfo {author} {\bibfnamefont {T.}~\bibnamefont {Loew}}, \bibinfo {author} {\bibfnamefont {J.}~\bibnamefont {Porras}}, \bibinfo {author} {\bibfnamefont {B.}~\bibnamefont {Keimer}}, \bibinfo {author} {\bibfnamefont {T.}~\bibnamefont {Shibauchi}},\ and\ \bibinfo {author} {\bibfnamefont {Y.}~\bibnamefont {Matsuda}},\ }\bibfield  {title} {\bibinfo {title} {{Thermodynamic evidence for a nematic phase transition at the onset of the pseudogap in YBa$_{2}$Cu$_{3}$O$_{y}$}},\ }\href {https://doi.org/10.1038/nphys4205} {\bibfield  {journal} {\bibinfo  {journal} {Nature physics}\ }\textbf {\bibinfo {volume} {13}},\
  \bibinfo {pages} {1074} (\bibinfo {year} {2017})}\BibitemShut {NoStop}%
\bibitem [{\citenamefont {Keimer}\ \emph {et~al.}(2015)\citenamefont {Keimer}, \citenamefont {Kivelson}, \citenamefont {Norman}, \citenamefont {Uchida},\ and\ \citenamefont {Zaanen}}]{Keimer2015-jp}%
  \BibitemOpen
  \bibfield  {author} {\bibinfo {author} {\bibfnamefont {B.}~\bibnamefont {Keimer}}, \bibinfo {author} {\bibfnamefont {S.~A.}\ \bibnamefont {Kivelson}}, \bibinfo {author} {\bibfnamefont {M.~R.}\ \bibnamefont {Norman}}, \bibinfo {author} {\bibfnamefont {S.}~\bibnamefont {Uchida}},\ and\ \bibinfo {author} {\bibfnamefont {J.}~\bibnamefont {Zaanen}},\ }\bibfield  {title} {\bibinfo {title} {{From quantum matter to high-temperature superconductivity in copper oxides}},\ }\href {https://doi.org/10.1038/nature14165} {\bibfield  {journal} {\bibinfo  {journal} {Nature}\ }\textbf {\bibinfo {volume} {518}},\ \bibinfo {pages} {179} (\bibinfo {year} {2015})}\BibitemShut {NoStop}%
\bibitem [{\citenamefont {Fradkin}\ \emph {et~al.}(2015)\citenamefont {Fradkin}, \citenamefont {Kivelson},\ and\ \citenamefont {Tranquada}}]{Fradkin2015-xz}%
  \BibitemOpen
  \bibfield  {author} {\bibinfo {author} {\bibfnamefont {E.}~\bibnamefont {Fradkin}}, \bibinfo {author} {\bibfnamefont {S.~A.}\ \bibnamefont {Kivelson}},\ and\ \bibinfo {author} {\bibfnamefont {J.~M.}\ \bibnamefont {Tranquada}},\ }\bibfield  {title} {\bibinfo {title} {{Colloquium: Theory of intertwined orders in high temperature superconductors}},\ }\href {https://doi.org/10.1103/RevModPhys.87.457} {\bibfield  {journal} {\bibinfo  {journal} {Reviews of modern physics}\ }\textbf {\bibinfo {volume} {87}},\ \bibinfo {pages} {457} (\bibinfo {year} {2015})}\BibitemShut {NoStop}%
\bibitem [{\citenamefont {Mai}\ \emph {et~al.}(2021)\citenamefont {Mai}, \citenamefont {Balduzzi}, \citenamefont {Johnston},\ and\ \citenamefont {Maier}}]{Mai2021-wv}%
  \BibitemOpen
  \bibfield  {author} {\bibinfo {author} {\bibfnamefont {P.}~\bibnamefont {Mai}}, \bibinfo {author} {\bibfnamefont {G.}~\bibnamefont {Balduzzi}}, \bibinfo {author} {\bibfnamefont {S.}~\bibnamefont {Johnston}},\ and\ \bibinfo {author} {\bibfnamefont {T.~A.}\ \bibnamefont {Maier}},\ }\bibfield  {title} {\bibinfo {title} {{Orbital structure of the effective pairing interaction in the high-temperature superconducting cuprates}},\ }\href {https://doi.org/10.1038/s41535-021-00326-5} {\bibfield  {journal} {\bibinfo  {journal} {npj Quantum Materials}\ }\textbf {\bibinfo {volume} {6}},\ \bibinfo {pages} {1} (\bibinfo {year} {2021})}\BibitemShut {NoStop}%
\bibitem [{\citenamefont {Zhang}\ and\ \citenamefont {Rice}(1988)}]{Zhang1988-dw}%
  \BibitemOpen
  \bibfield  {author} {\bibinfo {author} {\bibfnamefont {F.~C.}\ \bibnamefont {Zhang}}\ and\ \bibinfo {author} {\bibfnamefont {T.~M.}\ \bibnamefont {Rice}},\ }\bibfield  {title} {\bibinfo {title} {{Effective Hamiltonian for the superconducting Cu oxides}},\ }\href {https://doi.org/10.1103/physrevb.37.3759} {\bibfield  {journal} {\bibinfo  {journal} {Physical review. B, Condensed matter}\ }\textbf {\bibinfo {volume} {37}},\ \bibinfo {pages} {3759} (\bibinfo {year} {1988})}\BibitemShut {NoStop}%
\bibitem [{\citenamefont {Arovas}\ \emph {et~al.}(2022)\citenamefont {Arovas}, \citenamefont {Berg}, \citenamefont {Kivelson},\ and\ \citenamefont {Raghu}}]{Arovas2022-ua}%
  \BibitemOpen
  \bibfield  {author} {\bibinfo {author} {\bibfnamefont {D.~P.}\ \bibnamefont {Arovas}}, \bibinfo {author} {\bibfnamefont {E.}~\bibnamefont {Berg}}, \bibinfo {author} {\bibfnamefont {S.~A.}\ \bibnamefont {Kivelson}},\ and\ \bibinfo {author} {\bibfnamefont {S.}~\bibnamefont {Raghu}},\ }\bibfield  {title} {\bibinfo {title} {{The Hubbard model}},\ }\href {https://doi.org/10.1146/annurev-conmatphys-031620-102024} {\bibfield  {journal} {\bibinfo  {journal} {Annual review of condensed matter physics}\ }\textbf {\bibinfo {volume} {13}},\ \bibinfo {pages} {239} (\bibinfo {year} {2022})}\BibitemShut {NoStop}%
\bibitem [{noa(2013)}]{noauthor_2013-wp}%
  \BibitemOpen
  \bibfield  {title} {\bibinfo {title} {{The Hubbard model at half a century}},\ }\href {https://doi.org/10.1038/nphys2759} {\bibfield  {journal} {\bibinfo  {journal} {Nature physics}\ }\textbf {\bibinfo {volume} {9}},\ \bibinfo {pages} {523} (\bibinfo {year} {2013})}\BibitemShut {NoStop}%
\bibitem [{\citenamefont {Qin}\ \emph {et~al.}(2022)\citenamefont {Qin}, \citenamefont {Sch{\"a}fer}, \citenamefont {Andergassen}, \citenamefont {Corboz},\ and\ \citenamefont {Gull}}]{qin2022hubbard}%
  \BibitemOpen
  \bibfield  {author} {\bibinfo {author} {\bibfnamefont {M.}~\bibnamefont {Qin}}, \bibinfo {author} {\bibfnamefont {T.}~\bibnamefont {Sch{\"a}fer}}, \bibinfo {author} {\bibfnamefont {S.}~\bibnamefont {Andergassen}}, \bibinfo {author} {\bibfnamefont {P.}~\bibnamefont {Corboz}},\ and\ \bibinfo {author} {\bibfnamefont {E.}~\bibnamefont {Gull}},\ }\bibfield  {title} {\bibinfo {title} {The hubbard model: A computational perspective},\ }\href@noop {} {\bibfield  {journal} {\bibinfo  {journal} {Annual Review of Condensed Matter Physics}\ }\textbf {\bibinfo {volume} {13}},\ \bibinfo {pages} {275} (\bibinfo {year} {2022})}\BibitemShut {NoStop}%
\bibitem [{\citenamefont {{Simons Collaboration on the Many-Electron Problem}}\ \emph {et~al.}(2015)\citenamefont {{Simons Collaboration on the Many-Electron Problem}}, \citenamefont {LeBlanc}, \citenamefont {Antipov}, \citenamefont {Becca}, \citenamefont {Bulik}, \citenamefont {Chan}, \citenamefont {Chung}, \citenamefont {Deng}, \citenamefont {Ferrero}, \citenamefont {Henderson}, \citenamefont {Jiménez-Hoyos}, \citenamefont {Kozik}, \citenamefont {Liu}, \citenamefont {Millis}, \citenamefont {Prokof'ev}, \citenamefont {Qin}, \citenamefont {Scuseria}, \citenamefont {Shi}, \citenamefont {Svistunov}, \citenamefont {Tocchio}, \citenamefont {Tupitsyn}, \citenamefont {White}, \citenamefont {Zhang}, \citenamefont {Zheng}, \citenamefont {Zhu},\ and\ \citenamefont {Gull}}]{Simons2015-jb}%
  \BibitemOpen
  \bibfield  {author} {\bibinfo {author} {\bibnamefont {{Simons Collaboration on the Many-Electron Problem}}}, \bibinfo {author} {\bibfnamefont {J.~P.~F.}\ \bibnamefont {LeBlanc}}, \bibinfo {author} {\bibfnamefont {A.~E.}\ \bibnamefont {Antipov}}, \bibinfo {author} {\bibfnamefont {F.}~\bibnamefont {Becca}}, \bibinfo {author} {\bibfnamefont {I.~W.}\ \bibnamefont {Bulik}}, \bibinfo {author} {\bibfnamefont {G.~K.-L.}\ \bibnamefont {Chan}}, \bibinfo {author} {\bibfnamefont {C.-M.}\ \bibnamefont {Chung}}, \bibinfo {author} {\bibfnamefont {Y.}~\bibnamefont {Deng}}, \bibinfo {author} {\bibfnamefont {M.}~\bibnamefont {Ferrero}}, \bibinfo {author} {\bibfnamefont {T.~M.}\ \bibnamefont {Henderson}}, \bibinfo {author} {\bibfnamefont {C.~A.}\ \bibnamefont {Jiménez-Hoyos}}, \bibinfo {author} {\bibfnamefont {E.}~\bibnamefont {Kozik}}, \bibinfo {author} {\bibfnamefont {X.-W.}\ \bibnamefont {Liu}}, \bibinfo {author} {\bibfnamefont {A.~J.}\ \bibnamefont {Millis}}, \bibinfo {author} {\bibfnamefont {N.~V.}\ \bibnamefont
  {Prokof'ev}}, \bibinfo {author} {\bibfnamefont {M.}~\bibnamefont {Qin}}, \bibinfo {author} {\bibfnamefont {G.~E.}\ \bibnamefont {Scuseria}}, \bibinfo {author} {\bibfnamefont {H.}~\bibnamefont {Shi}}, \bibinfo {author} {\bibfnamefont {B.~V.}\ \bibnamefont {Svistunov}}, \bibinfo {author} {\bibfnamefont {L.~F.}\ \bibnamefont {Tocchio}}, \bibinfo {author} {\bibfnamefont {I.~S.}\ \bibnamefont {Tupitsyn}}, \bibinfo {author} {\bibfnamefont {S.~R.}\ \bibnamefont {White}}, \bibinfo {author} {\bibfnamefont {S.}~\bibnamefont {Zhang}}, \bibinfo {author} {\bibfnamefont {B.-X.}\ \bibnamefont {Zheng}}, \bibinfo {author} {\bibfnamefont {Z.}~\bibnamefont {Zhu}},\ and\ \bibinfo {author} {\bibfnamefont {E.}~\bibnamefont {Gull}},\ }\bibfield  {title} {\bibinfo {title} {{Solutions of the Two-Dimensional Hubbard Model: Benchmarks and Results from a Wide Range of Numerical Algorithms}},\ }\href {https://doi.org/10.1103/PhysRevX.5.041041} {\bibfield  {journal} {\bibinfo  {journal} {Physical Review X}\ }\textbf {\bibinfo {volume}
  {5}},\ \bibinfo {pages} {041041} (\bibinfo {year} {2015})}\BibitemShut {NoStop}%
\bibitem [{\citenamefont {Scalapino}(2012)}]{Scalapino2012-jr}%
  \BibitemOpen
  \bibfield  {author} {\bibinfo {author} {\bibfnamefont {D.~J.}\ \bibnamefont {Scalapino}},\ }\bibfield  {title} {\bibinfo {title} {{A common thread: The pairing interaction for unconventional superconductors}},\ }\href {https://doi.org/10.1103/RevModPhys.84.1383} {\bibfield  {journal} {\bibinfo  {journal} {Reviews of modern physics}\ }\textbf {\bibinfo {volume} {84}},\ \bibinfo {pages} {1383} (\bibinfo {year} {2012})}\BibitemShut {NoStop}%
\bibitem [{\citenamefont {Zheng}\ \emph {et~al.}(2017)\citenamefont {Zheng}, \citenamefont {Chung}, \citenamefont {Corboz}, \citenamefont {Ehlers}, \citenamefont {Qin}, \citenamefont {Noack}, \citenamefont {Shi}, \citenamefont {White}, \citenamefont {Zhang},\ and\ \citenamefont {Chan}}]{Zheng2017-of}%
  \BibitemOpen
  \bibfield  {author} {\bibinfo {author} {\bibfnamefont {B.-X.}\ \bibnamefont {Zheng}}, \bibinfo {author} {\bibfnamefont {C.-M.}\ \bibnamefont {Chung}}, \bibinfo {author} {\bibfnamefont {P.}~\bibnamefont {Corboz}}, \bibinfo {author} {\bibfnamefont {G.}~\bibnamefont {Ehlers}}, \bibinfo {author} {\bibfnamefont {M.-P.}\ \bibnamefont {Qin}}, \bibinfo {author} {\bibfnamefont {R.~M.}\ \bibnamefont {Noack}}, \bibinfo {author} {\bibfnamefont {H.}~\bibnamefont {Shi}}, \bibinfo {author} {\bibfnamefont {S.~R.}\ \bibnamefont {White}}, \bibinfo {author} {\bibfnamefont {S.}~\bibnamefont {Zhang}},\ and\ \bibinfo {author} {\bibfnamefont {G.~K.-L.}\ \bibnamefont {Chan}},\ }\bibfield  {title} {\bibinfo {title} {{Stripe order in the underdoped region of the two-dimensional Hubbard model}},\ }\href {https://doi.org/10.1126/science.aam7127} {\bibfield  {journal} {\bibinfo  {journal} {Science}\ }\textbf {\bibinfo {volume} {358}},\ \bibinfo {pages} {1155} (\bibinfo {year} {2017})}\BibitemShut {NoStop}%
\bibitem [{\citenamefont {{Simons Collaboration on the Many-Electron Problem}}\ \emph {et~al.}(2020)\citenamefont {{Simons Collaboration on the Many-Electron Problem}}, \citenamefont {Qin}, \citenamefont {Chung}, \citenamefont {Shi}, \citenamefont {Vitali}, \citenamefont {Hubig}, \citenamefont {Schollwöck}, \citenamefont {White},\ and\ \citenamefont {Zhang}}]{qin2020absence}%
  \BibitemOpen
  \bibfield  {author} {\bibinfo {author} {\bibnamefont {{Simons Collaboration on the Many-Electron Problem}}}, \bibinfo {author} {\bibfnamefont {M.}~\bibnamefont {Qin}}, \bibinfo {author} {\bibfnamefont {C.-M.}\ \bibnamefont {Chung}}, \bibinfo {author} {\bibfnamefont {H.}~\bibnamefont {Shi}}, \bibinfo {author} {\bibfnamefont {E.}~\bibnamefont {Vitali}}, \bibinfo {author} {\bibfnamefont {C.}~\bibnamefont {Hubig}}, \bibinfo {author} {\bibfnamefont {U.}~\bibnamefont {Schollwöck}}, \bibinfo {author} {\bibfnamefont {S.~R.}\ \bibnamefont {White}},\ and\ \bibinfo {author} {\bibfnamefont {S.}~\bibnamefont {Zhang}},\ }\bibfield  {title} {\bibinfo {title} {{Absence of Superconductivity in the Pure Two-Dimensional Hubbard Model}},\ }\href {https://doi.org/10.1103/PhysRevX.10.031016} {\bibfield  {journal} {\bibinfo  {journal} {Physical Review X}\ }\textbf {\bibinfo {volume} {10}},\ \bibinfo {pages} {031016} (\bibinfo {year} {2020})}\BibitemShut {NoStop}%
\bibitem [{\citenamefont {Xu}\ \emph {et~al.}(2024)\citenamefont {Xu}, \citenamefont {Chung}, \citenamefont {Qin}, \citenamefont {Schollwöck}, \citenamefont {White},\ and\ \citenamefont {Zhang}}]{Xu2024-wt}%
  \BibitemOpen
  \bibfield  {author} {\bibinfo {author} {\bibfnamefont {H.}~\bibnamefont {Xu}}, \bibinfo {author} {\bibfnamefont {C.-M.}\ \bibnamefont {Chung}}, \bibinfo {author} {\bibfnamefont {M.}~\bibnamefont {Qin}}, \bibinfo {author} {\bibfnamefont {U.}~\bibnamefont {Schollwöck}}, \bibinfo {author} {\bibfnamefont {S.~R.}\ \bibnamefont {White}},\ and\ \bibinfo {author} {\bibfnamefont {S.}~\bibnamefont {Zhang}},\ }\bibfield  {title} {\bibinfo {title} {{Coexistence of superconductivity with partially filled stripes in the Hubbard model}},\ }\href {https://doi.org/10.1126/science.adh7691} {\bibfield  {journal} {\bibinfo  {journal} {Science}\ }\textbf {\bibinfo {volume} {384}},\ \bibinfo {pages} {eadh7691} (\bibinfo {year} {2024})}\BibitemShut {NoStop}%
\bibitem [{\citenamefont {Himeda}\ \emph {et~al.}(2002)\citenamefont {Himeda}, \citenamefont {Kato},\ and\ \citenamefont {Ogata}}]{Himeda2002-mc}%
  \BibitemOpen
  \bibfield  {author} {\bibinfo {author} {\bibfnamefont {A.}~\bibnamefont {Himeda}}, \bibinfo {author} {\bibfnamefont {T.}~\bibnamefont {Kato}},\ and\ \bibinfo {author} {\bibfnamefont {M.}~\bibnamefont {Ogata}},\ }\bibfield  {title} {\bibinfo {title} {{Stripe states with spatially oscillating d-wave superconductivity in the two-dimensional t-t'-J model}},\ }\href {https://doi.org/10.1103/PhysRevLett.88.117001} {\bibfield  {journal} {\bibinfo  {journal} {Physical review letters}\ }\textbf {\bibinfo {volume} {88}},\ \bibinfo {pages} {117001} (\bibinfo {year} {2002})}\BibitemShut {NoStop}%
\bibitem [{\citenamefont {Jiang}\ and\ \citenamefont {Devereaux}(2019)}]{Jiang2019-km}%
  \BibitemOpen
  \bibfield  {author} {\bibinfo {author} {\bibfnamefont {H.-C.}\ \bibnamefont {Jiang}}\ and\ \bibinfo {author} {\bibfnamefont {T.~P.}\ \bibnamefont {Devereaux}},\ }\bibfield  {title} {\bibinfo {title} {{Superconductivity in the doped Hubbard model and its interplay with next-nearest hopping t'}},\ }\href {https://doi.org/10.1126/science.aal5304} {\bibfield  {journal} {\bibinfo  {journal} {Science}\ }\textbf {\bibinfo {volume} {365}},\ \bibinfo {pages} {1424} (\bibinfo {year} {2019})}\BibitemShut {NoStop}%
\bibitem [{\citenamefont {Corboz}\ \emph {et~al.}(2014)\citenamefont {Corboz}, \citenamefont {Rice},\ and\ \citenamefont {Troyer}}]{Corboz2014-kr}%
  \BibitemOpen
  \bibfield  {author} {\bibinfo {author} {\bibfnamefont {P.}~\bibnamefont {Corboz}}, \bibinfo {author} {\bibfnamefont {T.~M.}\ \bibnamefont {Rice}},\ and\ \bibinfo {author} {\bibfnamefont {M.}~\bibnamefont {Troyer}},\ }\bibfield  {title} {\bibinfo {title} {{Competing states in the t-J model: uniform D-wave state versus stripe state}},\ }\href {https://doi.org/10.1103/PhysRevLett.113.046402} {\bibfield  {journal} {\bibinfo  {journal} {Physical review letters}\ }\textbf {\bibinfo {volume} {113}},\ \bibinfo {pages} {046402} (\bibinfo {year} {2014})}\BibitemShut {NoStop}%
\bibitem [{\citenamefont {Zheng}\ and\ \citenamefont {Chan}(2016)}]{Zheng2016-xt}%
  \BibitemOpen
  \bibfield  {author} {\bibinfo {author} {\bibfnamefont {B.-X.}\ \bibnamefont {Zheng}}\ and\ \bibinfo {author} {\bibfnamefont {G.~K.-L.}\ \bibnamefont {Chan}},\ }\bibfield  {title} {\bibinfo {title} {{Ground-state phase diagram of the square lattice Hubbard model from density matrix embedding theory}},\ }\href {https://doi.org/10.1103/PhysRevB.93.035126} {\bibfield  {journal} {\bibinfo  {journal} {Physical review. B, Condensed matter}\ }\textbf {\bibinfo {volume} {93}},\ \bibinfo {pages} {035126} (\bibinfo {year} {2016})}\BibitemShut {NoStop}%
\bibitem [{\citenamefont {Chen}\ \emph {et~al.}(2021)\citenamefont {Chen}, \citenamefont {Wang}, \citenamefont {Rebec}, \citenamefont {Jia}, \citenamefont {Hashimoto}, \citenamefont {Lu}, \citenamefont {Moritz}, \citenamefont {Moore}, \citenamefont {Devereaux},\ and\ \citenamefont {Shen}}]{Chen2021-xx}%
  \BibitemOpen
  \bibfield  {author} {\bibinfo {author} {\bibfnamefont {Z.}~\bibnamefont {Chen}}, \bibinfo {author} {\bibfnamefont {Y.}~\bibnamefont {Wang}}, \bibinfo {author} {\bibfnamefont {S.~N.}\ \bibnamefont {Rebec}}, \bibinfo {author} {\bibfnamefont {T.}~\bibnamefont {Jia}}, \bibinfo {author} {\bibfnamefont {M.}~\bibnamefont {Hashimoto}}, \bibinfo {author} {\bibfnamefont {D.}~\bibnamefont {Lu}}, \bibinfo {author} {\bibfnamefont {B.}~\bibnamefont {Moritz}}, \bibinfo {author} {\bibfnamefont {R.~G.}\ \bibnamefont {Moore}}, \bibinfo {author} {\bibfnamefont {T.~P.}\ \bibnamefont {Devereaux}},\ and\ \bibinfo {author} {\bibfnamefont {Z.-X.}\ \bibnamefont {Shen}},\ }\bibfield  {title} {\bibinfo {title} {{Anomalously strong near-neighbor attraction in doped 1D cuprate chains}},\ }\href {https://doi.org/10.1126/science.abf5174} {\bibfield  {journal} {\bibinfo  {journal} {Science}\ }\textbf {\bibinfo {volume} {373}},\ \bibinfo {pages} {1235} (\bibinfo {year} {2021})}\BibitemShut {NoStop}%
\bibitem [{\citenamefont {Wang}\ \emph {et~al.}(2021)\citenamefont {Wang}, \citenamefont {Chen}, \citenamefont {Shi}, \citenamefont {Moritz}, \citenamefont {Shen},\ and\ \citenamefont {Devereaux}}]{Wang2021-vs}%
  \BibitemOpen
  \bibfield  {author} {\bibinfo {author} {\bibfnamefont {Y.}~\bibnamefont {Wang}}, \bibinfo {author} {\bibfnamefont {Z.}~\bibnamefont {Chen}}, \bibinfo {author} {\bibfnamefont {T.}~\bibnamefont {Shi}}, \bibinfo {author} {\bibfnamefont {B.}~\bibnamefont {Moritz}}, \bibinfo {author} {\bibfnamefont {Z.-X.}\ \bibnamefont {Shen}},\ and\ \bibinfo {author} {\bibfnamefont {T.~P.}\ \bibnamefont {Devereaux}},\ }\bibfield  {title} {\bibinfo {title} {{Phonon-Mediated Long-Range Attractive Interaction in One-Dimensional Cuprates}},\ }\href {https://doi.org/10.1103/PhysRevLett.127.197003} {\bibfield  {journal} {\bibinfo  {journal} {Physical review letters}\ }\textbf {\bibinfo {volume} {127}},\ \bibinfo {pages} {197003} (\bibinfo {year} {2021})}\BibitemShut {NoStop}%
\bibitem [{\citenamefont {Tang}\ \emph {et~al.}(2023)\citenamefont {Tang}, \citenamefont {Moritz}, \citenamefont {Peng}, \citenamefont {Shen},\ and\ \citenamefont {Devereaux}}]{Tang2023-aa}%
  \BibitemOpen
  \bibfield  {author} {\bibinfo {author} {\bibfnamefont {T.}~\bibnamefont {Tang}}, \bibinfo {author} {\bibfnamefont {B.}~\bibnamefont {Moritz}}, \bibinfo {author} {\bibfnamefont {C.}~\bibnamefont {Peng}}, \bibinfo {author} {\bibfnamefont {Z.-X.}\ \bibnamefont {Shen}},\ and\ \bibinfo {author} {\bibfnamefont {T.~P.}\ \bibnamefont {Devereaux}},\ }\bibfield  {title} {\bibinfo {title} {{Traces of electron-phonon coupling in one-dimensional cuprates}},\ }\href {https://doi.org/10.1038/s41467-023-38408-6} {\bibfield  {journal} {\bibinfo  {journal} {Nature communications}\ }\textbf {\bibinfo {volume} {14}},\ \bibinfo {pages} {3129} (\bibinfo {year} {2023})}\BibitemShut {NoStop}%
\bibitem [{\citenamefont {Peng}\ \emph {et~al.}(2023)\citenamefont {Peng}, \citenamefont {Wang}, \citenamefont {Wen}, \citenamefont {Lee}, \citenamefont {Devereaux},\ and\ \citenamefont {Jiang}}]{Peng2023-on}%
  \BibitemOpen
  \bibfield  {author} {\bibinfo {author} {\bibfnamefont {C.}~\bibnamefont {Peng}}, \bibinfo {author} {\bibfnamefont {Y.}~\bibnamefont {Wang}}, \bibinfo {author} {\bibfnamefont {J.}~\bibnamefont {Wen}}, \bibinfo {author} {\bibfnamefont {Y.~S.}\ \bibnamefont {Lee}}, \bibinfo {author} {\bibfnamefont {T.~P.}\ \bibnamefont {Devereaux}},\ and\ \bibinfo {author} {\bibfnamefont {H.-C.}\ \bibnamefont {Jiang}},\ }\bibfield  {title} {\bibinfo {title} {{Enhanced superconductivity by near-neighbor attraction in the doped extended Hubbard model}},\ }\href {https://doi.org/10.1103/PhysRevB.107.L201102} {\bibfield  {journal} {\bibinfo  {journal} {Physical review. B, Condensed matter}\ }\textbf {\bibinfo {volume} {107}},\ \bibinfo {pages} {L201102} (\bibinfo {year} {2023})}\BibitemShut {NoStop}%
\bibitem [{\citenamefont {Qu}\ \emph {et~al.}(2022)\citenamefont {Qu}, \citenamefont {Chen}, \citenamefont {Jiang}, \citenamefont {Wang},\ and\ \citenamefont {Li}}]{Qu2022-yv}%
  \BibitemOpen
  \bibfield  {author} {\bibinfo {author} {\bibfnamefont {D.-W.}\ \bibnamefont {Qu}}, \bibinfo {author} {\bibfnamefont {B.-B.}\ \bibnamefont {Chen}}, \bibinfo {author} {\bibfnamefont {H.-C.}\ \bibnamefont {Jiang}}, \bibinfo {author} {\bibfnamefont {Y.}~\bibnamefont {Wang}},\ and\ \bibinfo {author} {\bibfnamefont {W.}~\bibnamefont {Li}},\ }\bibfield  {title} {\bibinfo {title} {{Spin-triplet pairing induced by near-neighbor attraction in the extended Hubbard model for cuprate chain}},\ }\href {https://doi.org/10.1038/s42005-022-01030-x} {\bibfield  {journal} {\bibinfo  {journal} {Communications Physics}\ }\textbf {\bibinfo {volume} {5}},\ \bibinfo {pages} {1} (\bibinfo {year} {2022})}\BibitemShut {NoStop}%
\bibitem [{\citenamefont {Chen}\ \emph {et~al.}(2023)\citenamefont {Chen}, \citenamefont {Wang},\ and\ \citenamefont {Chen}}]{Chen2023-py}%
  \BibitemOpen
  \bibfield  {author} {\bibinfo {author} {\bibfnamefont {W.-C.}\ \bibnamefont {Chen}}, \bibinfo {author} {\bibfnamefont {Y.}~\bibnamefont {Wang}},\ and\ \bibinfo {author} {\bibfnamefont {C.-C.}\ \bibnamefont {Chen}},\ }\bibfield  {title} {\bibinfo {title} {{Superconducting phases of the square-lattice extended Hubbard model}},\ }\href {https://doi.org/10.1103/PhysRevB.108.064514} {\bibfield  {journal} {\bibinfo  {journal} {Physical review. B, Condensed matter}\ }\textbf {\bibinfo {volume} {108}},\ \bibinfo {pages} {064514} (\bibinfo {year} {2023})}\BibitemShut {NoStop}%
\bibitem [{\citenamefont {Gukelberger}\ \emph {et~al.}(2014)\citenamefont {Gukelberger}, \citenamefont {Kozik}, \citenamefont {Pollet}, \citenamefont {Prokof'ev}, \citenamefont {Sigrist}, \citenamefont {Svistunov},\ and\ \citenamefont {Troyer}}]{Gukelberger2014-sf}%
  \BibitemOpen
  \bibfield  {author} {\bibinfo {author} {\bibfnamefont {J.}~\bibnamefont {Gukelberger}}, \bibinfo {author} {\bibfnamefont {E.}~\bibnamefont {Kozik}}, \bibinfo {author} {\bibfnamefont {L.}~\bibnamefont {Pollet}}, \bibinfo {author} {\bibfnamefont {N.}~\bibnamefont {Prokof'ev}}, \bibinfo {author} {\bibfnamefont {M.}~\bibnamefont {Sigrist}}, \bibinfo {author} {\bibfnamefont {B.}~\bibnamefont {Svistunov}},\ and\ \bibinfo {author} {\bibfnamefont {M.}~\bibnamefont {Troyer}},\ }\bibfield  {title} {\bibinfo {title} {{p-Wave superfluidity by spin-nematic Fermi surface deformation}},\ }\href {https://doi.org/10.1103/PhysRevLett.113.195301} {\bibfield  {journal} {\bibinfo  {journal} {Physical review letters}\ }\textbf {\bibinfo {volume} {113}},\ \bibinfo {pages} {195301} (\bibinfo {year} {2014})}\BibitemShut {NoStop}%
\bibitem [{\citenamefont {Kallin}(2012)}]{Kallin2012-pl}%
  \BibitemOpen
  \bibfield  {author} {\bibinfo {author} {\bibfnamefont {C.}~\bibnamefont {Kallin}},\ }\bibfield  {title} {\bibinfo {title} {{Chiral p-wave order in Sr2RuO4}},\ }\href {https://doi.org/10.1088/0034-4885/75/4/042501} {\bibfield  {journal} {\bibinfo  {journal} {Reports on Progress in Physics}\ }\textbf {\bibinfo {volume} {75}},\ \bibinfo {pages} {042501} (\bibinfo {year} {2012})}\BibitemShut {NoStop}%
\bibitem [{\citenamefont {Lee}\ and\ \citenamefont {Wen}(2008)}]{Lee2008-kc}%
  \BibitemOpen
  \bibfield  {author} {\bibinfo {author} {\bibfnamefont {P.~A.}\ \bibnamefont {Lee}}\ and\ \bibinfo {author} {\bibfnamefont {X.-G.}\ \bibnamefont {Wen}},\ }\bibfield  {title} {\bibinfo {title} {{Spin-triplet $p$-wave pairing in a three-orbital model for iron pnictide superconductors}},\ }\href {https://doi.org/10.1103/PhysRevB.78.144517} {\bibfield  {journal} {\bibinfo  {journal} {Physical review. B, Condensed matter}\ }\textbf {\bibinfo {volume} {78}},\ \bibinfo {pages} {144517} (\bibinfo {year} {2008})}\BibitemShut {NoStop}%
\bibitem [{\citenamefont {Guo}\ and\ \citenamefont {Tajima}(2023)}]{Guo2023-vr}%
  \BibitemOpen
  \bibfield  {author} {\bibinfo {author} {\bibfnamefont {Y.}~\bibnamefont {Guo}}\ and\ \bibinfo {author} {\bibfnamefont {H.}~\bibnamefont {Tajima}},\ }\bibfield  {title} {\bibinfo {title} {{Competition between pairing and tripling in one-dimensional fermions with coexistent $s$- and $p$-wave interactions}},\ }\href {https://doi.org/10.1103/PhysRevB.107.024511} {\bibfield  {journal} {\bibinfo  {journal} {Physical review. B, Condensed matter}\ }\textbf {\bibinfo {volume} {107}},\ \bibinfo {pages} {024511} (\bibinfo {year} {2023})}\BibitemShut {NoStop}%
\bibitem [{\citenamefont {Mackenzie}\ and\ \citenamefont {Maeno}(2003)}]{Mackenzie2003-mn}%
  \BibitemOpen
  \bibfield  {author} {\bibinfo {author} {\bibfnamefont {A.~P.}\ \bibnamefont {Mackenzie}}\ and\ \bibinfo {author} {\bibfnamefont {Y.}~\bibnamefont {Maeno}},\ }\bibfield  {title} {\bibinfo {title} {The superconductivity of sr$_2$ruo$_4$ and the physics of spin-triplet pairing},\ }\href@noop {} {\bibfield  {journal} {\bibinfo  {journal} {Reviews of Modern Physics}\ }\textbf {\bibinfo {volume} {75}},\ \bibinfo {pages} {657} (\bibinfo {year} {2003})}\BibitemShut {NoStop}%
\bibitem [{\citenamefont {Kinjo}\ \emph {et~al.}(2022)\citenamefont {Kinjo}, \citenamefont {Manago}, \citenamefont {Kitagawa}, \citenamefont {Mao}, \citenamefont {Yonezawa}, \citenamefont {Maeno},\ and\ \citenamefont {Ishida}}]{Kinjo2022-vi}%
  \BibitemOpen
  \bibfield  {author} {\bibinfo {author} {\bibfnamefont {K.}~\bibnamefont {Kinjo}}, \bibinfo {author} {\bibfnamefont {M.}~\bibnamefont {Manago}}, \bibinfo {author} {\bibfnamefont {S.}~\bibnamefont {Kitagawa}}, \bibinfo {author} {\bibfnamefont {Z.~Q.}\ \bibnamefont {Mao}}, \bibinfo {author} {\bibfnamefont {S.}~\bibnamefont {Yonezawa}}, \bibinfo {author} {\bibfnamefont {Y.}~\bibnamefont {Maeno}},\ and\ \bibinfo {author} {\bibfnamefont {K.}~\bibnamefont {Ishida}},\ }\bibfield  {title} {\bibinfo {title} {{Superconducting spin smecticity evidencing the Fulde-Ferrell-Larkin-Ovchinnikov state in Sr$_2$RuO$_4$}},\ }\href {https://doi.org/10.1126/science.abb0332} {\bibfield  {journal} {\bibinfo  {journal} {Science}\ }\textbf {\bibinfo {volume} {376}},\ \bibinfo {pages} {397} (\bibinfo {year} {2022})}\BibitemShut {NoStop}%
\bibitem [{\citenamefont {Sauls}(1994)}]{Sauls1994-vs}%
  \BibitemOpen
  \bibfield  {author} {\bibinfo {author} {\bibfnamefont {J.~A.}\ \bibnamefont {Sauls}},\ }\bibfield  {title} {\bibinfo {title} {{The order parameter for the superconducting phases of UPt\textsubscript{3}}},\ }\href {https://doi.org/10.1080/00018739400101475} {\bibfield  {journal} {\bibinfo  {journal} {Advances in Physics}\ }\textbf {\bibinfo {volume} {43}},\ \bibinfo {pages} {113} (\bibinfo {year} {1994})}\BibitemShut {NoStop}%
\bibitem [{\citenamefont {Jiao}\ \emph {et~al.}(2020)\citenamefont {Jiao}, \citenamefont {Howard}, \citenamefont {Ran}, \citenamefont {Wang}, \citenamefont {Rodriguez}, \citenamefont {Sigrist}, \citenamefont {Wang}, \citenamefont {Butch},\ and\ \citenamefont {Madhavan}}]{Jiao2020-ff}%
  \BibitemOpen
  \bibfield  {author} {\bibinfo {author} {\bibfnamefont {L.}~\bibnamefont {Jiao}}, \bibinfo {author} {\bibfnamefont {S.}~\bibnamefont {Howard}}, \bibinfo {author} {\bibfnamefont {S.}~\bibnamefont {Ran}}, \bibinfo {author} {\bibfnamefont {Z.}~\bibnamefont {Wang}}, \bibinfo {author} {\bibfnamefont {J.~O.}\ \bibnamefont {Rodriguez}}, \bibinfo {author} {\bibfnamefont {M.}~\bibnamefont {Sigrist}}, \bibinfo {author} {\bibfnamefont {Z.}~\bibnamefont {Wang}}, \bibinfo {author} {\bibfnamefont {N.~P.}\ \bibnamefont {Butch}},\ and\ \bibinfo {author} {\bibfnamefont {V.}~\bibnamefont {Madhavan}},\ }\bibfield  {title} {\bibinfo {title} {{Chiral superconductivity in heavy-fermion metal UTe\textsubscript{2}}},\ }\href {https://doi.org/10.1038/s41586-020-2122-2} {\bibfield  {journal} {\bibinfo  {journal} {Nature}\ }\textbf {\bibinfo {volume} {579}},\ \bibinfo {pages} {523} (\bibinfo {year} {2020})}\BibitemShut {NoStop}%
\bibitem [{\citenamefont {Jia}\ \emph {et~al.}(2021)\citenamefont {Jia}, \citenamefont {Wang}, \citenamefont {Chiu}, \citenamefont {Song}, \citenamefont {Yu}, \citenamefont {Jäck}, \citenamefont {Lei}, \citenamefont {Klemenz}, \citenamefont {Cevallos}, \citenamefont {Onyszczak}, \citenamefont {Fishchenko}, \citenamefont {Liu}, \citenamefont {Farahi}, \citenamefont {Xie}, \citenamefont {Xu}, \citenamefont {Watanabe}, \citenamefont {Taniguchi}, \citenamefont {Bernevig}, \citenamefont {Cava}, \citenamefont {Schoop}, \citenamefont {Yazdani},\ and\ \citenamefont {Wu}}]{Jia2021-xw}%
  \BibitemOpen
  \bibfield  {author} {\bibinfo {author} {\bibfnamefont {Y.}~\bibnamefont {Jia}}, \bibinfo {author} {\bibfnamefont {P.}~\bibnamefont {Wang}}, \bibinfo {author} {\bibfnamefont {C.-L.}\ \bibnamefont {Chiu}}, \bibinfo {author} {\bibfnamefont {Z.}~\bibnamefont {Song}}, \bibinfo {author} {\bibfnamefont {G.}~\bibnamefont {Yu}}, \bibinfo {author} {\bibfnamefont {B.}~\bibnamefont {Jäck}}, \bibinfo {author} {\bibfnamefont {S.}~\bibnamefont {Lei}}, \bibinfo {author} {\bibfnamefont {S.}~\bibnamefont {Klemenz}}, \bibinfo {author} {\bibfnamefont {F.~A.}\ \bibnamefont {Cevallos}}, \bibinfo {author} {\bibfnamefont {M.}~\bibnamefont {Onyszczak}}, \bibinfo {author} {\bibfnamefont {N.}~\bibnamefont {Fishchenko}}, \bibinfo {author} {\bibfnamefont {X.}~\bibnamefont {Liu}}, \bibinfo {author} {\bibfnamefont {G.}~\bibnamefont {Farahi}}, \bibinfo {author} {\bibfnamefont {F.}~\bibnamefont {Xie}}, \bibinfo {author} {\bibfnamefont {Y.}~\bibnamefont {Xu}}, \bibinfo {author} {\bibfnamefont {K.}~\bibnamefont {Watanabe}}, \bibinfo
  {author} {\bibfnamefont {T.}~\bibnamefont {Taniguchi}}, \bibinfo {author} {\bibfnamefont {B.~A.}\ \bibnamefont {Bernevig}}, \bibinfo {author} {\bibfnamefont {R.~J.}\ \bibnamefont {Cava}}, \bibinfo {author} {\bibfnamefont {L.~M.}\ \bibnamefont {Schoop}}, \bibinfo {author} {\bibfnamefont {A.}~\bibnamefont {Yazdani}},\ and\ \bibinfo {author} {\bibfnamefont {S.}~\bibnamefont {Wu}},\ }\bibfield  {title} {\bibinfo {title} {{Evidence for a monolayer excitonic insulator}},\ }\href {https://doi.org/10.1038/s41567-021-01422-w} {\bibfield  {journal} {\bibinfo  {journal} {Nature physics}\ }\textbf {\bibinfo {volume} {18}},\ \bibinfo {pages} {87} (\bibinfo {year} {2021})}\BibitemShut {NoStop}%
\bibitem [{\citenamefont {Matano}\ \emph {et~al.}(2016)\citenamefont {Matano}, \citenamefont {Kriener}, \citenamefont {Segawa}, \citenamefont {Ando},\ and\ \citenamefont {Zheng}}]{Matano_undated-ie}%
  \BibitemOpen
  \bibfield  {author} {\bibinfo {author} {\bibfnamefont {K.}~\bibnamefont {Matano}}, \bibinfo {author} {\bibfnamefont {M.}~\bibnamefont {Kriener}}, \bibinfo {author} {\bibfnamefont {K.}~\bibnamefont {Segawa}}, \bibinfo {author} {\bibfnamefont {Y.}~\bibnamefont {Ando}},\ and\ \bibinfo {author} {\bibfnamefont {G.-q.}\ \bibnamefont {Zheng}},\ }\bibfield  {title} {\bibinfo {title} {Spin-rotation symmetry breaking in the superconducting state of $cu_xbi_2se_3$},\ }\href {https://doi.org/10.1038/nphys3781} {\bibfield  {journal} {\bibinfo  {journal} {Nature Physics}\ }\textbf {\bibinfo {volume} {12}},\ \bibinfo {pages} {852} (\bibinfo {year} {2016})}\BibitemShut {NoStop}%
\bibitem [{\citenamefont {Liu}\ \emph {et~al.}(2023)\citenamefont {Liu}, \citenamefont {Wei}, \citenamefont {He}, \citenamefont {Zhang}, \citenamefont {Wang},\ and\ \citenamefont {Wang}}]{Liu2023-vc}%
  \BibitemOpen
  \bibfield  {author} {\bibinfo {author} {\bibfnamefont {Y.}~\bibnamefont {Liu}}, \bibinfo {author} {\bibfnamefont {T.}~\bibnamefont {Wei}}, \bibinfo {author} {\bibfnamefont {G.}~\bibnamefont {He}}, \bibinfo {author} {\bibfnamefont {Y.}~\bibnamefont {Zhang}}, \bibinfo {author} {\bibfnamefont {Z.}~\bibnamefont {Wang}},\ and\ \bibinfo {author} {\bibfnamefont {J.}~\bibnamefont {Wang}},\ }\bibfield  {title} {\bibinfo {title} {{Pair density wave state in a monolayer high-Tc iron-based superconductor}},\ }\href {https://doi.org/10.1038/s41586-023-06072-x} {\bibfield  {journal} {\bibinfo  {journal} {Nature}\ }\textbf {\bibinfo {volume} {618}},\ \bibinfo {pages} {934} (\bibinfo {year} {2023})}\BibitemShut {NoStop}%
\bibitem [{\citenamefont {Sarma}\ \emph {et~al.}(2015)\citenamefont {Sarma}, \citenamefont {Freedman},\ and\ \citenamefont {Nayak}}]{Sarma2015-jq}%
  \BibitemOpen
  \bibfield  {author} {\bibinfo {author} {\bibfnamefont {S.~D.}\ \bibnamefont {Sarma}}, \bibinfo {author} {\bibfnamefont {M.}~\bibnamefont {Freedman}},\ and\ \bibinfo {author} {\bibfnamefont {C.}~\bibnamefont {Nayak}},\ }\bibfield  {title} {\bibinfo {title} {{Majorana zero modes and topological quantum computation}},\ }\href {https://doi.org/10.1038/npjqi.2015.1} {\bibfield  {journal} {\bibinfo  {journal} {npj Quantum Information}\ }\textbf {\bibinfo {volume} {1}},\ \bibinfo {pages} {1} (\bibinfo {year} {2015})}\BibitemShut {NoStop}%
\bibitem [{\citenamefont {Crépel}\ and\ \citenamefont {Fu}(2022)}]{Crepel2022-be}%
  \BibitemOpen
  \bibfield  {author} {\bibinfo {author} {\bibfnamefont {V.}~\bibnamefont {Crépel}}\ and\ \bibinfo {author} {\bibfnamefont {L.}~\bibnamefont {Fu}},\ }\bibfield  {title} {\bibinfo {title} {{Spin-triplet superconductivity from excitonic effect in doped insulators}},\ }\href {https://doi.org/10.1073/pnas.2117735119} {\bibfield  {journal} {\bibinfo  {journal} {Proceedings of the National Academy of Sciences of the United States of America}\ }\textbf {\bibinfo {volume} {119}},\ \bibinfo {pages} {e2117735119} (\bibinfo {year} {2022})}\BibitemShut {NoStop}%
\bibitem [{\citenamefont {Zhang}\ \emph {et~al.}(1995)\citenamefont {Zhang}, \citenamefont {Carlson},\ and\ \citenamefont {Gubernatis}}]{Zhang1995-hn}%
  \BibitemOpen
  \bibfield  {author} {\bibinfo {author} {\bibfnamefont {S.}~\bibnamefont {Zhang}}, \bibinfo {author} {\bibfnamefont {J.}~\bibnamefont {Carlson}},\ and\ \bibinfo {author} {\bibfnamefont {J.~E.}\ \bibnamefont {Gubernatis}},\ }\bibfield  {title} {\bibinfo {title} {{Constrained path quantum Monte Carlo method for fermion ground states}},\ }\href {https://doi.org/10.1103/PhysRevLett.74.3652} {\bibfield  {journal} {\bibinfo  {journal} {Physical review letters}\ }\textbf {\bibinfo {volume} {74}},\ \bibinfo {pages} {3652} (\bibinfo {year} {1995})}\BibitemShut {NoStop}%
\bibitem [{\citenamefont {Zhang}\ \emph {et~al.}(1997)\citenamefont {Zhang}, \citenamefont {Carlson},\ and\ \citenamefont {Gubernatis}}]{Zhang1997-mr}%
  \BibitemOpen
  \bibfield  {author} {\bibinfo {author} {\bibfnamefont {S.}~\bibnamefont {Zhang}}, \bibinfo {author} {\bibfnamefont {J.}~\bibnamefont {Carlson}},\ and\ \bibinfo {author} {\bibfnamefont {J.~E.}\ \bibnamefont {Gubernatis}},\ }\bibfield  {title} {\bibinfo {title} {{Constrained path Monte Carlo method for fermion ground states}},\ }\href {https://doi.org/10.1103/PhysRevB.55.7464} {\bibfield  {journal} {\bibinfo  {journal} {Physical review. B, Condensed matter}\ }\textbf {\bibinfo {volume} {55}},\ \bibinfo {pages} {7464} (\bibinfo {year} {1997})}\BibitemShut {NoStop}%
\bibitem [{Cao()}]{Cao2024V-supp}%
  \BibitemOpen
  \href@noop {} {\bibinfo {title} {{See Supplemental Material, which includes Ref. \cite{Liu2021-mj,romer2020pairing,Loh1990-iq}, for details on the method and additional simulation results.}}}\BibitemShut {Stop}%
\bibitem [{\citenamefont {Andersen}\ \emph {et~al.}(1995)\citenamefont {Andersen}, \citenamefont {Liechtenstein}, \citenamefont {Jepsen},\ and\ \citenamefont {Paulsen}}]{Andersen1995-pi}%
  \BibitemOpen
  \bibfield  {author} {\bibinfo {author} {\bibfnamefont {O.~K.}\ \bibnamefont {Andersen}}, \bibinfo {author} {\bibfnamefont {A.~I.}\ \bibnamefont {Liechtenstein}}, \bibinfo {author} {\bibfnamefont {O.}~\bibnamefont {Jepsen}},\ and\ \bibinfo {author} {\bibfnamefont {F.}~\bibnamefont {Paulsen}},\ }\bibfield  {title} {\bibinfo {title} {{LDA energy bands, low-energy hamiltonians, $t^{\prime}$, $t^{\prime\prime}$, $t_{\bot}(k)$, and $J_{\bot}$}},\ }\href {https://doi.org/10.1016/0022-3697(95)00269-3} {\bibfield  {journal} {\bibinfo  {journal} {The Journal of physics and chemistry of solids}\ }\textbf {\bibinfo {volume} {56}},\ \bibinfo {pages} {1573} (\bibinfo {year} {1995})}\BibitemShut {NoStop}%
\bibitem [{\citenamefont {Hirayama}\ \emph {et~al.}(2018)\citenamefont {Hirayama}, \citenamefont {Yamaji}, \citenamefont {Misawa},\ and\ \citenamefont {Imada}}]{Hirayama2018-tm}%
  \BibitemOpen
  \bibfield  {author} {\bibinfo {author} {\bibfnamefont {M.}~\bibnamefont {Hirayama}}, \bibinfo {author} {\bibfnamefont {Y.}~\bibnamefont {Yamaji}}, \bibinfo {author} {\bibfnamefont {T.}~\bibnamefont {Misawa}},\ and\ \bibinfo {author} {\bibfnamefont {M.}~\bibnamefont {Imada}},\ }\bibfield  {title} {\bibinfo {title} {{Ab initio effective Hamiltonians for cuprate superconductors}},\ }\href {https://doi.org/10.1103/PhysRevB.98.134501} {\bibfield  {journal} {\bibinfo  {journal} {Physical review. B, Condensed matter}\ }\textbf {\bibinfo {volume} {98}},\ \bibinfo {pages} {134501} (\bibinfo {year} {2018})}\BibitemShut {NoStop}%
\bibitem [{\citenamefont {Dahm}\ \emph {et~al.}(2009)\citenamefont {Dahm}, \citenamefont {Hinkov}, \citenamefont {Borisenko}, \citenamefont {Kordyuk}, \citenamefont {Zabolotnyy}, \citenamefont {Fink}, \citenamefont {Büchner}, \citenamefont {Scalapino}, \citenamefont {Hanke},\ and\ \citenamefont {Keimer}}]{Dahm2009-wg}%
  \BibitemOpen
  \bibfield  {author} {\bibinfo {author} {\bibfnamefont {T.}~\bibnamefont {Dahm}}, \bibinfo {author} {\bibfnamefont {V.}~\bibnamefont {Hinkov}}, \bibinfo {author} {\bibfnamefont {S.~V.}\ \bibnamefont {Borisenko}}, \bibinfo {author} {\bibfnamefont {A.~A.}\ \bibnamefont {Kordyuk}}, \bibinfo {author} {\bibfnamefont {V.~B.}\ \bibnamefont {Zabolotnyy}}, \bibinfo {author} {\bibfnamefont {J.}~\bibnamefont {Fink}}, \bibinfo {author} {\bibfnamefont {B.}~\bibnamefont {Büchner}}, \bibinfo {author} {\bibfnamefont {D.~J.}\ \bibnamefont {Scalapino}}, \bibinfo {author} {\bibfnamefont {W.}~\bibnamefont {Hanke}},\ and\ \bibinfo {author} {\bibfnamefont {B.}~\bibnamefont {Keimer}},\ }\bibfield  {title} {\bibinfo {title} {{Strength of the spin-fluctuation-mediated pairing interaction in a high-temperature superconductor}},\ }\href {https://doi.org/10.1038/nphys1180} {\bibfield  {journal} {\bibinfo  {journal} {Nature physics}\ }\textbf {\bibinfo {volume} {5}},\ \bibinfo {pages} {217} (\bibinfo {year} {2009})}\BibitemShut
  {NoStop}%
\bibitem [{\citenamefont {Dong}\ \emph {et~al.}(2022)\citenamefont {Dong}, \citenamefont {Del~Re}, \citenamefont {Toschi},\ and\ \citenamefont {Gull}}]{Dong2022-xh}%
  \BibitemOpen
  \bibfield  {author} {\bibinfo {author} {\bibfnamefont {X.}~\bibnamefont {Dong}}, \bibinfo {author} {\bibfnamefont {L.}~\bibnamefont {Del~Re}}, \bibinfo {author} {\bibfnamefont {A.}~\bibnamefont {Toschi}},\ and\ \bibinfo {author} {\bibfnamefont {E.}~\bibnamefont {Gull}},\ }\bibfield  {title} {\bibinfo {title} {{Mechanism of superconductivity in the Hubbard model at intermediate interaction strength}},\ }\href {https://doi.org/10.1073/pnas.2205048119} {\bibfield  {journal} {\bibinfo  {journal} {Proceedings of the National Academy of Sciences of the United States of America}\ }\textbf {\bibinfo {volume} {119}},\ \bibinfo {pages} {e2205048119} (\bibinfo {year} {2022})}\BibitemShut {NoStop}%
\bibitem [{\citenamefont {Otsuka}\ \emph {et~al.}(2020)\citenamefont {Otsuka}, \citenamefont {Seki}, \citenamefont {Sorella},\ and\ \citenamefont {Yunoki}}]{Otsuka2020-wc}%
  \BibitemOpen
  \bibfield  {author} {\bibinfo {author} {\bibfnamefont {Y.}~\bibnamefont {Otsuka}}, \bibinfo {author} {\bibfnamefont {K.}~\bibnamefont {Seki}}, \bibinfo {author} {\bibfnamefont {S.}~\bibnamefont {Sorella}},\ and\ \bibinfo {author} {\bibfnamefont {S.}~\bibnamefont {Yunoki}},\ }\bibfield  {title} {\bibinfo {title} {{Dirac electrons in the square-lattice Hubbard model with a d -wave pairing field: The chiral Heisenberg universality class revisited}},\ }\href {https://doi.org/10.1103/physrevb.102.235105} {\bibfield  {journal} {\bibinfo  {journal} {Physical review. B}\ }\textbf {\bibinfo {volume} {102}},\ \bibinfo {pages} {235105} (\bibinfo {year} {2020})}\BibitemShut {NoStop}%
\bibitem [{\citenamefont {Luther}\ and\ \citenamefont {Peschel}(1975)}]{Luther1975-jd}%
  \BibitemOpen
  \bibfield  {author} {\bibinfo {author} {\bibfnamefont {A.}~\bibnamefont {Luther}}\ and\ \bibinfo {author} {\bibfnamefont {I.}~\bibnamefont {Peschel}},\ }\bibfield  {title} {\bibinfo {title} {{Calculation of critical exponents in two dimensions from quantum field theory in one dimension}},\ }\href {https://doi.org/10.1103/physrevb.12.3908} {\bibfield  {journal} {\bibinfo  {journal} {Physical review b}\ }\textbf {\bibinfo {volume} {12}},\ \bibinfo {pages} {3908} (\bibinfo {year} {1975})}\BibitemShut {NoStop}%
\bibitem [{\citenamefont {Schulz}(1990)}]{Schulz1990-tb}%
  \BibitemOpen
  \bibfield  {author} {\bibinfo {author} {\bibfnamefont {H.~J.}\ \bibnamefont {Schulz}},\ }\bibfield  {title} {\bibinfo {title} {{Correlation exponents and the metal-insulator transition in the one-dimensional Hubbard model}},\ }\href {https://doi.org/10.1103/PhysRevLett.64.2831} {\bibfield  {journal} {\bibinfo  {journal} {Physical review letters}\ }\textbf {\bibinfo {volume} {64}},\ \bibinfo {pages} {2831} (\bibinfo {year} {1990})}\BibitemShut {NoStop}%
\bibitem [{\citenamefont {Das}(2020)}]{Das2020-yp}%
  \BibitemOpen
  \bibfield  {author} {\bibinfo {author} {\bibfnamefont {T.}~\bibnamefont {Das}},\ }\bibfield  {title} {\bibinfo {title} {{Nodeless superconducting gap induced by odd parity pair density wave in underdoped cuprates}},\ }\href {https://doi.org/10.1016/j.aop.2020.168251} {\bibfield  {journal} {\bibinfo  {journal} {Annals of physics}\ }\textbf {\bibinfo {volume} {420}},\ \bibinfo {pages} {168251} (\bibinfo {year} {2020})}\BibitemShut {NoStop}%
\bibitem [{\citenamefont {Lu}\ \emph {et~al.}(2014)\citenamefont {Lu}, \citenamefont {Xiang},\ and\ \citenamefont {Lee}}]{Lu2014-bn}%
  \BibitemOpen
  \bibfield  {author} {\bibinfo {author} {\bibfnamefont {Y.-M.}\ \bibnamefont {Lu}}, \bibinfo {author} {\bibfnamefont {T.}~\bibnamefont {Xiang}},\ and\ \bibinfo {author} {\bibfnamefont {D.-H.}\ \bibnamefont {Lee}},\ }\bibfield  {title} {\bibinfo {title} {{Underdoped superconducting cuprates as topological superconductors}},\ }\href {https://doi.org/10.1038/nphys3021} {\bibfield  {journal} {\bibinfo  {journal} {Nature physics}\ }\textbf {\bibinfo {volume} {10}},\ \bibinfo {pages} {634} (\bibinfo {year} {2014})}\BibitemShut {NoStop}%
\bibitem [{\citenamefont {Liu}\ \emph {et~al.}(2021)\citenamefont {Liu}, \citenamefont {Yang}, \citenamefont {Li}, \citenamefont {Ying}, \citenamefont {Yang}, \citenamefont {Sun},\ and\ \citenamefont {Li}}]{Liu2021-mj}%
  \BibitemOpen
  \bibfield  {author} {\bibinfo {author} {\bibfnamefont {K.}~\bibnamefont {Liu}}, \bibinfo {author} {\bibfnamefont {S.}~\bibnamefont {Yang}}, \bibinfo {author} {\bibfnamefont {W.}~\bibnamefont {Li}}, \bibinfo {author} {\bibfnamefont {T.}~\bibnamefont {Ying}}, \bibinfo {author} {\bibfnamefont {J.}~\bibnamefont {Yang}}, \bibinfo {author} {\bibfnamefont {X.}~\bibnamefont {Sun}},\ and\ \bibinfo {author} {\bibfnamefont {X.}~\bibnamefont {Li}},\ }\bibfield  {title} {\bibinfo {title} {{The pairing symmetries in the two-dimensional Hubbard model}},\ }\href {https://doi.org/10.1016/j.physleta.2021.127153} {\bibfield  {journal} {\bibinfo  {journal} {Physics letters. A}\ }\textbf {\bibinfo {volume} {392}},\ \bibinfo {pages} {127153} (\bibinfo {year} {2021})}\BibitemShut {NoStop}%
\bibitem [{\citenamefont {R{\o}mer}\ \emph {et~al.}(2020)\citenamefont {R{\o}mer}, \citenamefont {Maier}, \citenamefont {Kreisel}, \citenamefont {Eremin}, \citenamefont {Hirschfeld},\ and\ \citenamefont {Andersen}}]{romer2020pairing}%
  \BibitemOpen
  \bibfield  {author} {\bibinfo {author} {\bibfnamefont {A.~T.}\ \bibnamefont {R{\o}mer}}, \bibinfo {author} {\bibfnamefont {T.~A.}\ \bibnamefont {Maier}}, \bibinfo {author} {\bibfnamefont {A.}~\bibnamefont {Kreisel}}, \bibinfo {author} {\bibfnamefont {I.}~\bibnamefont {Eremin}}, \bibinfo {author} {\bibfnamefont {P.}~\bibnamefont {Hirschfeld}},\ and\ \bibinfo {author} {\bibfnamefont {B.~M.}\ \bibnamefont {Andersen}},\ }\bibfield  {title} {\bibinfo {title} {Pairing in the two-dimensional hubbard model from weak to strong coupling},\ }\href@noop {} {\bibfield  {journal} {\bibinfo  {journal} {Physical Review Research}\ }\textbf {\bibinfo {volume} {2}},\ \bibinfo {pages} {013108} (\bibinfo {year} {2020})}\BibitemShut {NoStop}%
\bibitem [{\citenamefont {Loh}\ \emph {et~al.}(1990)\citenamefont {Loh}, \citenamefont {Gubernatis}, \citenamefont {Scalettar}, \citenamefont {White}, \citenamefont {Scalapino},\ and\ \citenamefont {Sugar}}]{Loh1990-iq}%
  \BibitemOpen
  \bibfield  {author} {\bibinfo {author} {\bibfnamefont {E.~Y.}\ \bibnamefont {Loh}}, \bibinfo {author} {\bibfnamefont {J.~E.}\ \bibnamefont {Gubernatis}}, \bibinfo {author} {\bibfnamefont {R.~T.}\ \bibnamefont {Scalettar}}, \bibinfo {author} {\bibfnamefont {S.~R.}\ \bibnamefont {White}}, \bibinfo {author} {\bibfnamefont {D.~J.}\ \bibnamefont {Scalapino}},\ and\ \bibinfo {author} {\bibfnamefont {R.~L.}\ \bibnamefont {Sugar}},\ }\bibfield  {title} {\bibinfo {title} {{Sign problem in the numerical simulation of many-electron systems}},\ }\href {https://doi.org/10.1103/PhysRevB.41.9301} {\bibfield  {journal} {\bibinfo  {journal} {Physical review. B, Condensed matter}\ }\textbf {\bibinfo {volume} {41}},\ \bibinfo {pages} {9301} (\bibinfo {year} {1990})}\BibitemShut {NoStop}%
\end{thebibliography}%


\begin{thebibliography}{5}%
\makeatletter
\providecommand \@ifxundefined [1]{%
 \@ifx{#1\undefined}
}%
\providecommand \@ifnum [1]{%
 \ifnum #1\expandafter \@firstoftwo
 \else \expandafter \@secondoftwo
 \fi
}%
\providecommand \@ifx [1]{%
 \ifx #1\expandafter \@firstoftwo
 \else \expandafter \@secondoftwo
 \fi
}%
\providecommand \natexlab [1]{#1}%
\providecommand \enquote  [1]{``#1''}%
\providecommand \bibnamefont  [1]{#1}%
\providecommand \bibfnamefont [1]{#1}%
\providecommand \citenamefont [1]{#1}%
\providecommand \href@noop [0]{\@secondoftwo}%
\providecommand \href [0]{\begingroup \@sanitize@url \@href}%
\providecommand \@href[1]{\@@startlink{#1}\@@href}%
\providecommand \@@href[1]{\endgroup#1\@@endlink}%
\providecommand \@sanitize@url [0]{\catcode `\\12\catcode `\$12\catcode `\&12\catcode `\#12\catcode `\^12\catcode `\_12\catcode `\%12\relax}%
\providecommand \@@startlink[1]{}%
\providecommand \@@endlink[0]{}%
\providecommand \url  [0]{\begingroup\@sanitize@url \@url }%
\providecommand \@url [1]{\endgroup\@href {#1}{\urlprefix }}%
\providecommand \urlprefix  [0]{URL }%
\providecommand \Eprint [0]{\href }%
\providecommand \doibase [0]{https://doi.org/}%
\providecommand \selectlanguage [0]{\@gobble}%
\providecommand \bibinfo  [0]{\@secondoftwo}%
\providecommand \bibfield  [0]{\@secondoftwo}%
\providecommand \translation [1]{[#1]}%
\providecommand \BibitemOpen [0]{}%
\providecommand \bibitemStop [0]{}%
\providecommand \bibitemNoStop [0]{.\EOS\space}%
\providecommand \EOS [0]{\spacefactor3000\relax}%
\providecommand \BibitemShut  [1]{\csname bibitem#1\endcsname}%
\let\auto@bib@innerbib\@empty
%</preamble>
\bibitem [{\citenamefont {Liu}\ \emph {et~al.}(2021)\citenamefont {Liu}, \citenamefont {Yang}, \citenamefont {Li}, \citenamefont {Ying}, \citenamefont {Yang}, \citenamefont {Sun},\ and\ \citenamefont {Li}}]{Liu2021-mj}%
  \BibitemOpen
  \bibfield  {author} {\bibinfo {author} {\bibfnamefont {K.}~\bibnamefont {Liu}}, \bibinfo {author} {\bibfnamefont {S.}~\bibnamefont {Yang}}, \bibinfo {author} {\bibfnamefont {W.}~\bibnamefont {Li}}, \bibinfo {author} {\bibfnamefont {T.}~\bibnamefont {Ying}}, \bibinfo {author} {\bibfnamefont {J.}~\bibnamefont {Yang}}, \bibinfo {author} {\bibfnamefont {X.}~\bibnamefont {Sun}},\ and\ \bibinfo {author} {\bibfnamefont {X.}~\bibnamefont {Li}},\ }\bibfield  {title} {\bibinfo {title} {{The pairing symmetries in the two-dimensional Hubbard model}},\ }\href {https://doi.org/10.1016/j.physleta.2021.127153} {\bibfield  {journal} {\bibinfo  {journal} {Physics letters. A}\ }\textbf {\bibinfo {volume} {392}},\ \bibinfo {pages} {127153} (\bibinfo {year} {2021})}\BibitemShut {NoStop}%
\bibitem [{\citenamefont {R{\o}mer}\ \emph {et~al.}(2020)\citenamefont {R{\o}mer}, \citenamefont {Maier}, \citenamefont {Kreisel}, \citenamefont {Eremin}, \citenamefont {Hirschfeld},\ and\ \citenamefont {Andersen}}]{romer2020pairing}%
  \BibitemOpen
  \bibfield  {author} {\bibinfo {author} {\bibfnamefont {A.~T.}\ \bibnamefont {R{\o}mer}}, \bibinfo {author} {\bibfnamefont {T.~A.}\ \bibnamefont {Maier}}, \bibinfo {author} {\bibfnamefont {A.}~\bibnamefont {Kreisel}}, \bibinfo {author} {\bibfnamefont {I.}~\bibnamefont {Eremin}}, \bibinfo {author} {\bibfnamefont {P.}~\bibnamefont {Hirschfeld}},\ and\ \bibinfo {author} {\bibfnamefont {B.~M.}\ \bibnamefont {Andersen}},\ }\bibfield  {title} {\bibinfo {title} {Pairing in the two-dimensional hubbard model from weak to strong coupling},\ }\href@noop {} {\bibfield  {journal} {\bibinfo  {journal} {Physical Review Research}\ }\textbf {\bibinfo {volume} {2}},\ \bibinfo {pages} {013108} (\bibinfo {year} {2020})}\BibitemShut {NoStop}%
\bibitem [{\citenamefont {Zhang}\ \emph {et~al.}(1995)\citenamefont {Zhang}, \citenamefont {Carlson},\ and\ \citenamefont {Gubernatis}}]{Zhang1995-hn}%
  \BibitemOpen
  \bibfield  {author} {\bibinfo {author} {\bibfnamefont {S.}~\bibnamefont {Zhang}}, \bibinfo {author} {\bibfnamefont {J.}~\bibnamefont {Carlson}},\ and\ \bibinfo {author} {\bibfnamefont {J.~E.}\ \bibnamefont {Gubernatis}},\ }\bibfield  {title} {\bibinfo {title} {{Constrained path quantum Monte Carlo method for fermion ground states}},\ }\href {https://doi.org/10.1103/PhysRevLett.74.3652} {\bibfield  {journal} {\bibinfo  {journal} {Physical review letters}\ }\textbf {\bibinfo {volume} {74}},\ \bibinfo {pages} {3652} (\bibinfo {year} {1995})}\BibitemShut {NoStop}%
\bibitem [{\citenamefont {Zhang}\ \emph {et~al.}(1997)\citenamefont {Zhang}, \citenamefont {Carlson},\ and\ \citenamefont {Gubernatis}}]{Zhang1997-mr}%
  \BibitemOpen
  \bibfield  {author} {\bibinfo {author} {\bibfnamefont {S.}~\bibnamefont {Zhang}}, \bibinfo {author} {\bibfnamefont {J.}~\bibnamefont {Carlson}},\ and\ \bibinfo {author} {\bibfnamefont {J.~E.}\ \bibnamefont {Gubernatis}},\ }\bibfield  {title} {\bibinfo {title} {{Constrained path Monte Carlo method for fermion ground states}},\ }\href {https://doi.org/10.1103/PhysRevB.55.7464} {\bibfield  {journal} {\bibinfo  {journal} {Physical review. B, Condensed matter}\ }\textbf {\bibinfo {volume} {55}},\ \bibinfo {pages} {7464} (\bibinfo {year} {1997})}\BibitemShut {NoStop}%
\bibitem [{\citenamefont {Loh}\ \emph {et~al.}(1990)\citenamefont {Loh}, \citenamefont {Gubernatis}, \citenamefont {Scalettar}, \citenamefont {White}, \citenamefont {Scalapino},\ and\ \citenamefont {Sugar}}]{Loh1990-iq}%
  \BibitemOpen
  \bibfield  {author} {\bibinfo {author} {\bibfnamefont {E.~Y.}\ \bibnamefont {Loh}}, \bibinfo {author} {\bibfnamefont {J.~E.}\ \bibnamefont {Gubernatis}}, \bibinfo {author} {\bibfnamefont {R.~T.}\ \bibnamefont {Scalettar}}, \bibinfo {author} {\bibfnamefont {S.~R.}\ \bibnamefont {White}}, \bibinfo {author} {\bibfnamefont {D.~J.}\ \bibnamefont {Scalapino}},\ and\ \bibinfo {author} {\bibfnamefont {R.~L.}\ \bibnamefont {Sugar}},\ }\bibfield  {title} {\bibinfo {title} {{Sign problem in the numerical simulation of many-electron systems}},\ }\href {https://doi.org/10.1103/PhysRevB.41.9301} {\bibfield  {journal} {\bibinfo  {journal} {Physical review. B, Condensed matter}\ }\textbf {\bibinfo {volume} {41}},\ \bibinfo {pages} {9301} (\bibinfo {year} {1990})}\BibitemShut {NoStop}%
\end{thebibliography}%
\end{document}

\title{Supplemental Material for ``Dominant $p$-wave pairing induced by near-neighbor attraction in the square-lattice extended Hubbard model"}

\author{Zhangkai Cao}
\thanks{These authors contributed equally.}
\affiliation{School of Science, Harbin Institute of Technology, Shenzhen, 518055, China}

\author{Jianyu Li}
\thanks{These authors contributed equally.}
\affiliation{School of Science, Harbin Institute of Technology, Shenzhen, 518055, China}
\affiliation{Shenzhen Key Laboratory of Advanced Functional Carbon Materials Research and Comprehensive Application, Shenzhen 518055, China.}
 
\author{Jiahao Su}
\affiliation{School of Science, Harbin Institute of Technology, Shenzhen, 518055, China}
\affiliation{Shenzhen Key Laboratory of Advanced Functional Carbon Materials Research and Comprehensive Application, Shenzhen 518055, China.}

\author{Tao Ying}
\email{taoying86@hit.edu.cn}
\affiliation{School of Physics, Harbin Institute of Technology, Harbin 150001, China}

\author{Ho-Kin Tang}
\email{denghaojian@hit.edu.cn}
\affiliation{School of Science, Harbin Institute of Technology, Shenzhen, 518055, China}
\affiliation{Shenzhen Key Laboratory of Advanced Functional Carbon Materials Research and Comprehensive Application, Shenzhen 518055, China.}

\date{\today} 
\maketitle 

In this Supplemental Material, we present additional computational results. In Sec.~S1, we have defined various correlation functions. In Sec.~S2, we provide the effective d-wave and p-wave pairing correlation function in momentum space with the change of $U$, $V$ and $\delta$. In Sec.~S3, we give a phase diagram of the dominant pairing channel in $U-n$ space when $V$=0. In Sec.~S4, we give a brief introduction to constraint path quantum Monte Carlo (CPQMC).

% In this Supplemental Material, we present additional definitions and computational results. In Sec.~S1, we have defined various correlation functions. In Sec.~S2, we give two possible processes of the two-boson $d^{(1)}_{x^2-y^2}$-wave pairing correlations on the NN sites. In Sec.~S3, we mainly introduce the transition of $s$-SF phase to CPBM phase induced by anisotropy in different $t^{\prime}$. In Sec.~S4, we discuss the determination of the CPBM phase. In Sec.~S5, we discuss the boson correlation of various symmetry. In Sec.~S6, we investigated the competition between $d$-wave boson correlation when $t<t^{\prime}$. In Sec.~S7, we show the incommensurate density wave disappears in large $t^{\prime}$. In Sec.~S8, we give a brief introduction to Constraint path quantum Monte Carlo.

\section{Correlation function}

\begin{figure}[]
    \centering 
    \includegraphics[width=0.6\linewidth]{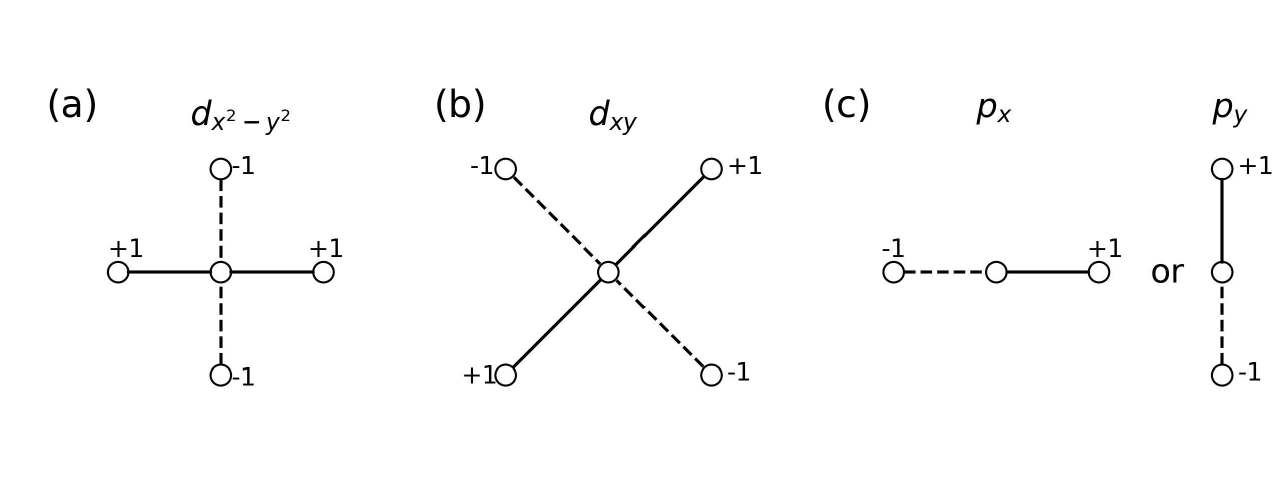}
    \caption{(Color online) Phase of the $d_{x^2-y^2},\ d_{xy},\ p\ (p_x$ or $p_y$) pairing symmetry.  }
    \label{figS1}
\end{figure}

We have defined the effective pair momentum distribution function
\begin{equation}
N^{\rm eff}_{\mathrm \zeta-pair}({\bf k}) = (1/N)\sum_{i,j} \mbox{exp}[i{\bf k}({\bf r}_i-{\bf r}_j)]C^{\rm eff}_{\mathrm \zeta-pair}(i,j),
\label{nspdtkpair}
\end{equation}
where $\zeta =  s,\ d_{x^2-y^2},\ d_{xy},\ p$. 
The effective real-space correlation of the on-site $s$-wave pairing operator $C^{\rm eff}_{s-pair}(i,j) = \langle {\Delta}_{s}^{\dagger}(i) {\Delta}_{s}(j) \rangle - G^{\uparrow}_{i,j}G^{\downarrow}_{i,j}$, where $\Delta_s^\dagger(i)=c^\dagger_{i\uparrow}c^\dagger_{i\downarrow}$ is the on-site $s$-wave pairing operator and $G^{\uparrow}_{i,j}G^{\downarrow}_{i,j}$ is the uncorrelated pairing structure factor. $G^{\sigma}_{i,j}= \langle c_{i\sigma}c^\dagger_{j\sigma} \rangle$ is the Green's function. The effective real-space correlation of the NN sites $d_{x^2-y^2}$-wave pairing operator $C^{\rm eff}_{d_{x^2-y^2}-pair}(i,j) = \sum_{\delta_{\zeta},\delta'_{\zeta}}\langle {\Delta}_{d_{x^2-y^2}}^{\dagger}(i,i+\delta_{\zeta}) {\Delta}_{d_{x^2-y^2}}(j,j+\delta'_{\zeta}) \rangle - G^{\uparrow}_{i,j}G^{\downarrow}_{i+\delta_{\zeta},j+\delta'_{\zeta}})$, where $\Delta_{d_{x^2-y^2}}^\dagger(i)=c^\dagger_{i\uparrow}(c^\dagger_{i+x \downarrow}-c^\dagger_{i+y \downarrow}+c^\dagger_{i-x \downarrow}-c^\dagger_{i-y \downarrow})$ and $\delta^{(')}_{\zeta}$ are NN sites in the $d_{x^2-y^2}$-wave pairing. The effective real-space correlation of the NNN sites $d_{xy}$-wave pairing operator $C^{\rm eff}_{d_{xy}-pair}(i,j) = \sum_{\delta_{\zeta},\delta'_{\zeta}}\langle {\Delta}_{d_{xy}}^{\dagger}(i,i+\delta_{\zeta}) {\Delta}_{d_{xy}}(j,j+\delta'_{\zeta}) \rangle - G^{\uparrow}_{i,j}G^{\downarrow}_{i+\delta_{\zeta},j+\delta'_{\zeta}})$, where $\Delta_{d_{xy}}^\dagger(i)=c^\dagger_{i\uparrow}(c^\dagger_{i+x+y \downarrow}-c^\dagger_{i-x+y \downarrow}+c^\dagger_{i-x-y \downarrow}-c^\dagger_{i+x-y \downarrow})$ and $\delta^{(')}_{\zeta}$ are the NNN sites in the $d_{xy}$-wave pairing.
The effective real-space correlation of the NN sites $p$-wave pairing operator $C^{\rm eff}_{p-pair}(i,j) = \sum_{\delta_{\zeta},\delta'_{\zeta}}\langle {\Delta}_{p}^{\dagger}(i,i+\delta_{\zeta}) {\Delta}_{p}(j,j+\delta'_{\zeta}) \rangle - G^{\sigma}_{i,j}G^{\sigma}_{i+\delta_{\zeta},j+\delta'_{\zeta}})$, in which $\delta^{(')}_{\zeta}$ are the NN sites in the $p$-wave pairing. $\Delta_{p}^\dagger(i)=c^\dagger_{i\sigma}(c^\dagger_{i+l \sigma}-c^\dagger_{i-l \sigma})$, with spin $\sigma=\uparrow, \downarrow$ representing the $\uparrow \uparrow$ and $\downarrow \downarrow$ pairing, and $l = x,\ y$ corresponds to the symmetries of $p_x$ and $p_y$, respectively. Considering  the structure of square lattice, these different pairing symmetries have the following form factor, as shown in Fig.\ \ref{figS1}.
The effective on-site $s$-wave pairing exists only in the weak $U$ and large $V$ regions. 
The values of the effective $s$-wave are small and do not dominate the system, so it was not included it in the main text.

% To study the density correlations, we defined the charge structure factor and spin structure factor in particle-hole channel, 
% \begin{equation}
% N_{\mathrm c}({\bf k}) = (1/N)\sum_{i,j} \mbox{exp}[i{\bf k}({\bf r}_i-{\bf r}_j)]\, \langle {n}_i {n}_j \rangle,
% \label{nskpair}
% \end{equation}
% \begin{equation}
% N_{\mathrm s}({\bf k}) = (1/N)\sum_{i,j} \mbox{exp}[i{\bf k}({\bf r}_i-{\bf r}_j)]\, \langle {\bf S}_i \cdot{\bf S}_j \rangle,
% \label{nskpair}
% \end{equation}
% Here, the density number operator is defined as
% ${n}_i = \sum_{\sigma} c^\dagger_{i \sigma} c_{i \sigma}$ and ${\bf S}_i$ is the spin operator at site $i$.

% % The correlation function of different pairing mode in real space are defined as $C_{\mathrm \zeta-pair} = \langle {\Delta}_{\zeta}^{\dagger}(i) {\Delta}_{\zeta}(j) \rangle$.
% The correlation function of charge density wave and spin density wave in real space are defined as $C_{\rm CDW} = \langle {n}_i {n}_j \rangle$ and $C_{\rm SDW} = \langle {\bf S}_i \cdot{\bf S}_j \rangle$.

\section{The effective $d$-wave and $p$-wave pairing in momentum space evolve with the change of $U$, $V$ and $\delta$}

\begin{figure}[b!]
    \centering 
    \includegraphics[width=1\linewidth]{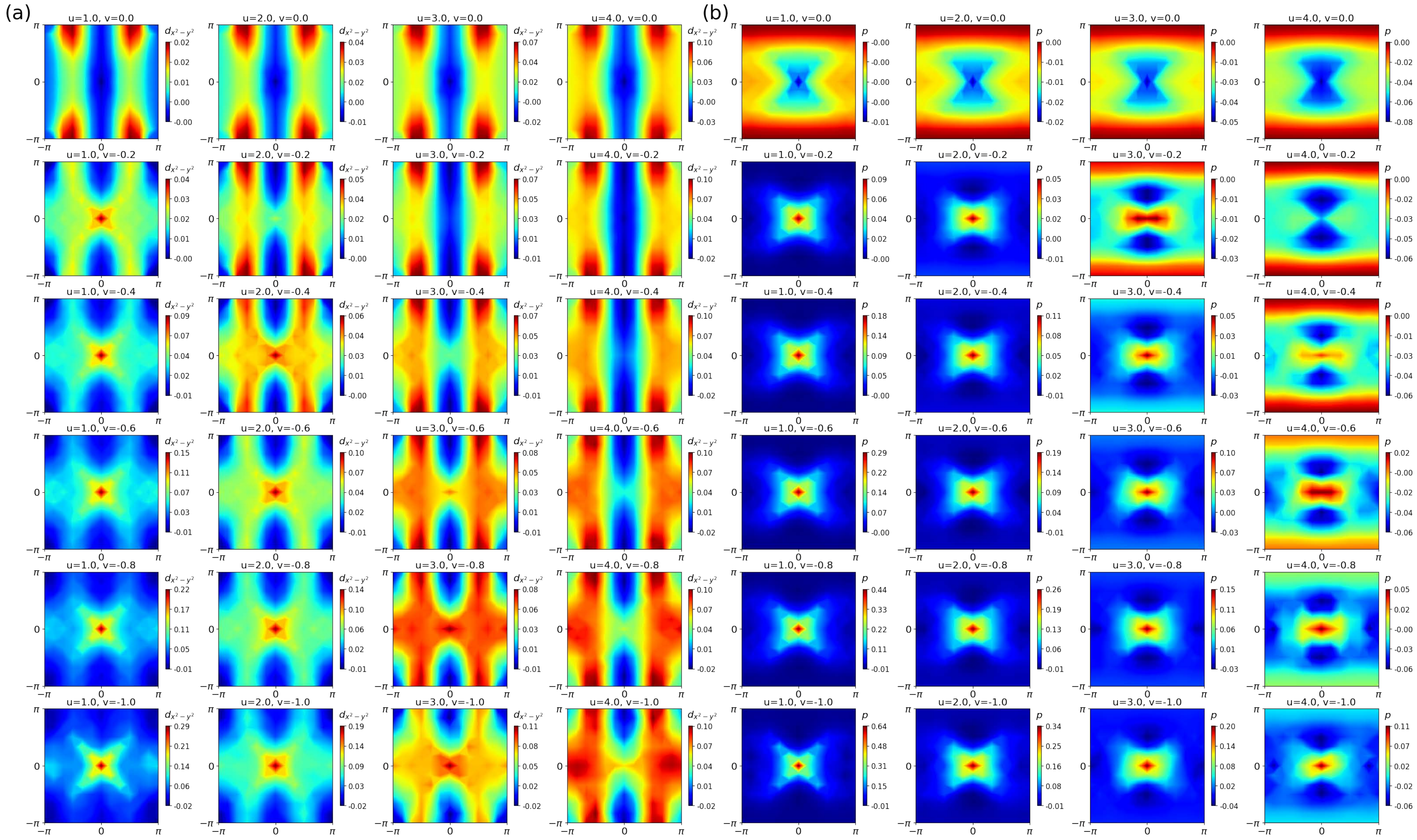}
    \caption{(Color online) (a) CPQMC simulation result of effective $d$-wave pair momentum distribution function $N^{\rm eff}_{d_{x^2-y^2}-pair}({\bf k})$ at $\delta$ = 0.125 with $U=1, 2, 3, 4$ and $V= 0.0, -0.2, -0.4, -0.6, -0.8, -1.0$ on 12 $\times$ 12 lattice. (b) CPQMC simulation result of effective $p$-wave pair momentum distribution function $N^{\rm eff}_{p-pair}({\bf k})$ at $\delta$ = 0.125 with $U=1, 2, 3, 4$ and $V= 0.0, -0.2, -0.4, -0.6, -0.8, -1.0$ on 12 $\times$ 12 lattice.   }
    \label{figS2}
\end{figure}

This section we analyze the characteristic of effective $d$-wave and $p$-wave pairing correlation function in momentum space evolving with the change of $U$, $V$ and $\delta$. In Fig.\ \ref{figS2}(a), we show the effective $d$-wave pairing correlation function $N^{\rm eff}_{d_{x^2-y^2}-pair}({\bf k})$ at $\delta$= 0.125 with $U=1, 2, 3, 4$ and $V= 0.0, -0.2, -0.4, -0.6, -0.8, -1.0$. We can compare the phase diagram (Fig. 1(b)) to the dominant region of the $d_{x^2-y^2}$-wave pairing, $d$-wave pairing correlation function exhibits a singular nonzero condensation structure, and its peak occurs near nonzero momentum point $Q=(2\pi/3,\ \pi)$, which demonstrate the presence of the $d$-wave PDW phase. However, in the dominant region of the $p$-wave pairing, $d$-wave pairing gradually evolves into a $Q=(0,\ 0)$ condensation, exhibiting a significantly weaker strength compared to the $p$-wave pairing. 
In Fig.\ \ref{figS2}(b), we show the effective $p$-wave pairing correlation function $N^{\rm eff}_{\mathrm p-pair}({\bf k})$ in momentum space at $\delta =$ 0.125 with $U=1, 2, 3, 4$ and $V= 0.0, -0.2, -0.4, -0.6, -0.8, -1.0$. It clearly that the $p$-wave pairing strength increases with the increase of $V$ and suppressed by the increase of $U$. Noteworthy, when the maximum value in momentum space is less than 0, it indicates the absence of $p$-wave pairing. In the dominant region of the $p$-wave pairing, the $p$-wave pairing always condenses at point $Q=(0,\ 0)$. 

\begin{figure}[!]
    \centering 
    \includegraphics[width=0.6\linewidth]{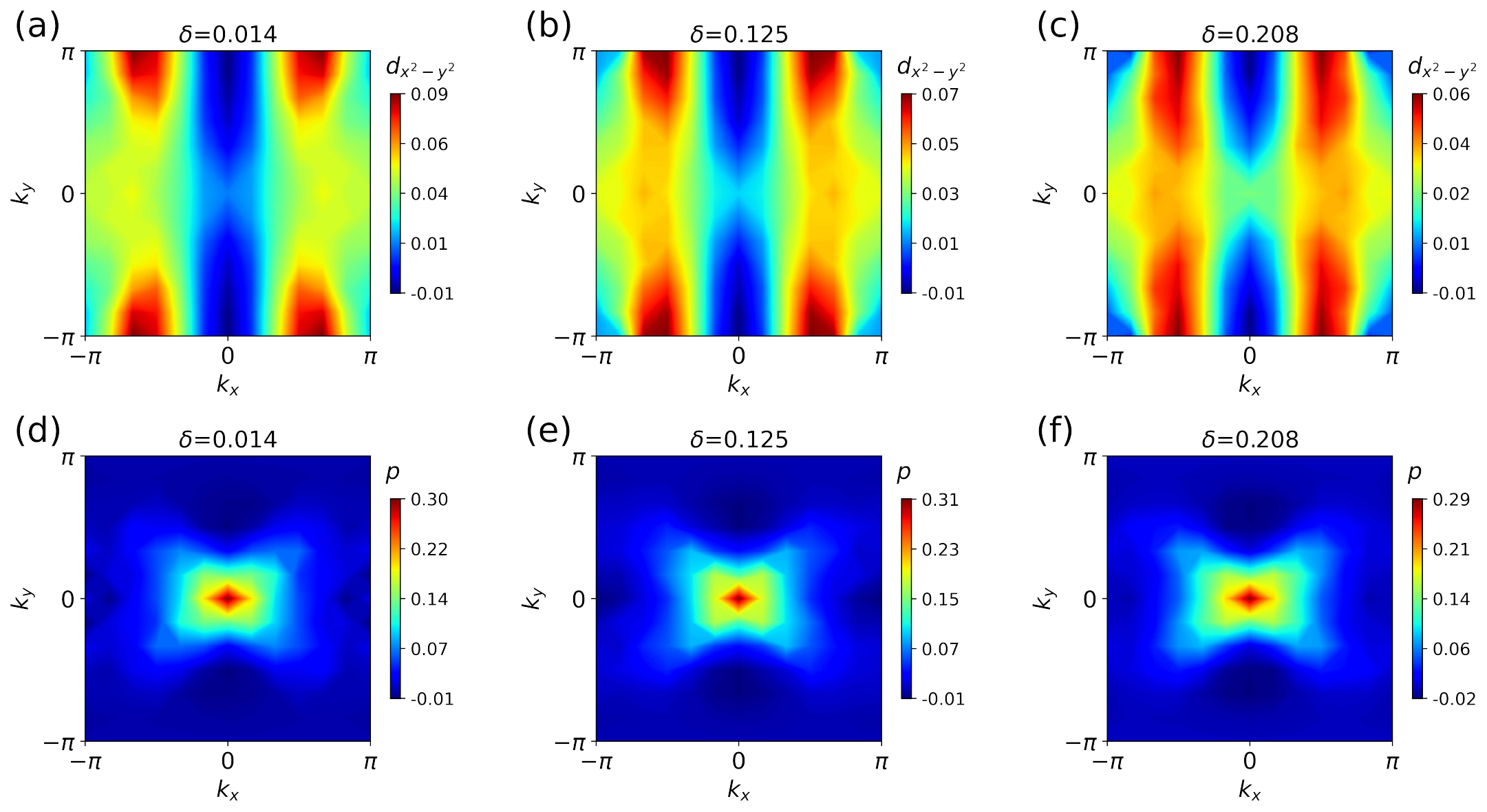}
    \caption{(Color online) (a)-(c) CPQMC simulation result of effective $d$-wave pair momentum distribution function $N^{\rm eff}_{d_{x^2-y^2}-pair}({\bf k})$ at $\delta$ = 0.014, 0.125, 0.208 with $U= 3$ and $V= -0.3$ on 12 $\times$ 12 lattice. (d)-(f) CPQMC simulation result of effective $p$-wave pair momentum distribution function $N^{\rm eff}_{p-pair}({\bf k})$ at $\delta$= 0.014, 0.125, 0.208 with $U= 2$ and $V= -0.9$ on 12 $\times$ 12 lattice. } 
    \label{figS3}
\end{figure}

We also present how the effective $d$-wave and $p$-wave pairings evolve with doping. As shown in Fig.\ \ref{figS3}(a)-(c), with doping increases, we can see the condensation point becomes slightly dispersed and the intensity of the $d$-wave pairing decreases, which may be related to the suppression of the $d$-wave by doping.
At the same time, the $p$-wave pairing has always been zero momentum condensation, and its intensity has remained basically unchanged when evolve with the change of $\delta$. This indicates that as $\delta$ increases, the expansion of the $p$-wave pairing dominant region is mainly due to the decrease of $d$-wave pairing.

In the dominant region of $d$-wave pairing, the position of its condensation point will not change significantly, but with the change of $U$, $V$ and $\delta$, the sharpness of its nonzero condensation point will change slightly. In the dominant region of the $p$-wave pairing, it always condenses at point $Q=(0,\ 0)$, which supporting the formation of Cooper-pair triplets with zero center-of-mass momentum and $p$-wave symmetry.

% \section{The influence of $V$ on SDW order near half filling}

% \begin{figure}[b!]
%     \centering 
%     \includegraphics[width=0.8\linewidth]{figS4.pdf}
%     \caption{(Color online) (a)-(c) Lattice size effect for spin structure factor $N_{S}({\bf k})$ with $U$ = 4.0, 3.5, 3.0 at near half filling $\delta$ $\sim$ 0.014 for various values of $V$, fitted using second-order polynomials in $1/L$. }
%     \label{figS4}
% \end{figure}

% \zhang{In Fig.\ \ref{figS4}(a)-(c), we shows the CPQMC results for the finite-size extrapolation at near half filling $\delta$ $\sim$ 0.014 with $U$ = 4.0, 3.5, 3.0 and $V$ = 0.0, -1.0. The SDW order appears at near half filling $\delta$ $\sim$ 0.014 when $U$ = 4.0, the extrapolated value is greater than 0. On the contrary, as $U$ decreases, the extrapolated value is less than 0 and SDW order does not exist. Especially, when $U$ = 3.5, there is no SDW order when $V=-1.0$, but SDW order still exists when $V$=0. This also indicates that the presence of $V$ is not conducive to the formation of SDW order, but this effect is still relatively weak.
% So overall, the system needed a larger critical $U$ to form the SDW order in the presence of an attractive $V$.

% }

\section{$U-n$ phase diagram}

\begin{figure}[!]
    \centering 
    \includegraphics[width=0.5\linewidth]{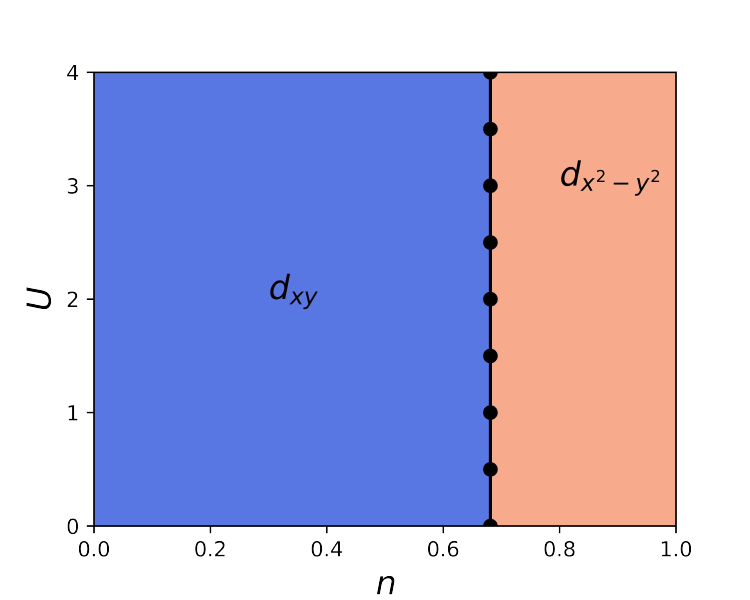}
    \caption{(Color online) The zero-temperature schematic phase diagram of the dominant pairing channel in $U-n$ space when $V$=0.  }
    \label{figS4}
\end{figure}

% We have defined the $d_{xy}$-wave pair momentum distribution function
% \begin{equation}
% N_{\mathrm \zeta-pair}({\bf k}) = (1/N)\sum_{i,j} \mbox{exp}[i{\bf k}({\bf r}_i-{\bf r}_j)]\, \langle {\Delta}_{\zeta}^{\dagger}(i) {\Delta}_{\zeta}(j) \rangle,
% \label{nspdtkpair}
% \end{equation}
% where $\zeta = d_{xy}$. $N_{\mathrm d_{xy}-pair}({\bf k})$ is the next NN sites d-wave pair momentum distribution function and $\Delta_{d_{xy}}^\dagger(i)=c^\dagger_{i\uparrow}(c^\dagger_{i+x+y \downarrow}-c^\dagger_{i-x+y \downarrow}+c^\dagger_{i-x-y \downarrow}-c^\dagger_{i+x-y \downarrow})$.

% Many previous studies of the Hubbard model have demonstrated that superconductivity can emerge from purely repulsive electron interactions \cite{Liu2021-mj,Jiang2022-yh,Sorella2021-nf,romer2020pairing}. Despite its simple form, however, the Hubbard model cannot be solved analytically in 2D system. So numerical algorithms were developed rapidly in recent years, such as the density matrix renormalization group (DMRG) \cite{Jiang2022-yh}, the variational auxiliary field quantum Monte Carlo (VAFQMC) \cite{Sorella2021-nf}, the dynamic cluster approximation (DCA) and the quantum Monte Carlo (QMC) algorithm \cite{romer2020pairing}, and determinant quantum Monte Carlo (DQMC) \cite{Liu2021-mj}. Till now, it was generally accepted that the $d_{x^2-y^2}$-wave is the dominant pairing channel in cuprate superconductors. 
In this section, we mainly explore the dominant pairing channels in different interaction strength and doping regimes when $V$= 0. 
At zero temperature, we summarize the phase diagram of the square-lattice extended Hubbard model in Fig.\ \ref{figS4}, where includes the $d_{x^2-y^2}$-wave and $d_{xy}$-wave pairing phases in $U-n$ space when $V$= 0. At $n=0.681-1$, the $d_{x^2-y^2}$-wave pairing phase dominates the system when there is no attraction $V$, while for small electron densities $n<0.681$, the dominant pairing channel is the $d_{xy}$-wave. This is a schematic phase diagram drawn based on 12 $\times$ 12 lattice, which will become more accurate as the system size increase.
% With the increase of interaction $U$, the boundary between $d_{x^2-y^2}$-wave and $d_{xy}$-wave shifts a little bit to lower electron density. 
Our results agree qualitatively with the dynamic cluster approximation (DCA) and determinant quantum Monte Carlo (DQMC) studies \cite{Liu2021-mj,romer2020pairing}, but there are differences in details. The phase boundary between the $d_{x^2-y^2}$-wave and $d_{xy}$-wave pairing is $n\sim 0.6$ in their study, but in our study, this boundary is $n\sim 0.7$. This is mainly because our CPQMC method is study the ground state of the system at zero temperature. Furthermore, we did not observe the presence of $p$-wave pairing at low electron density when $V$ = 0.

\section{Constraint path quantum Monte Carlo}

The CPQMC method is a quantum Monte Carlo method with a constraint path approximation~\cite{Zhang1995-hn,Zhang1997-mr}. CPQMC method prevents the infamous sign problem \cite{Loh1990-iq} encountered in the DQMC method when dealing with systems that are far from half filling. In CPQMC, the ground state wave function $\psi^{(n)}$ is represented by a finite ensemble of Slater determinants, i.e.,  
$\left|\psi^{(n)}\right\rangle \propto \sum_k\left|\phi_k^{(n)}\right\rangle$,
where $k$ is the index of the Slater determinants, and $n$ is the number of iteration. The overall normalization factor of the wave function has been omitted here. The propagation of the Slater determinants dominates the computational time, as follows
\begin{equation}
\left|\phi_k^{(n+1)}\right\rangle \leftarrow \int d \vec{x} P(\vec{x}) B(\vec{x})\left|\phi_k^{(n)}\right\rangle .
\label{propagation}
\end{equation}
where $\vec{x}$ is the auxiliary-field configuration, that we select according to the probability distribution function $P(\vec{x})$. The propagation includes the matrix multiplication of the propagator $B(\vec{x})$ and $\phi_k^{(n)}$. 
After a series of equilibrium steps, the walkers are the Monte Carlo samples of the ground state wave function $\phi^{(0)}$ and ground-state properties can be measured. 

The random walk formulation suffers from the sign problem because of the fundamental symmetry existing between the fermion ground state $\left|\psi_0\right\rangle$ and its negative $-\left|\psi_0\right\rangle $. 
In more general cases, walkers can cross $\mathcal{N}$ in their propagation by $e^{-\Delta \tau H}$
whose bounding surface $\mathcal{N}$ is defined by $\left\langle\psi_0 \mid \phi\right\rangle=0$ and is in general $unknown$.
Once a random walker reaches $\mathcal{N}$, it will make no further contribution to the representation of the ground state since
\begin{equation}
\left\langle\psi_0 \mid \phi\right\rangle=0 \Rightarrow\left\langle\psi_0\left|e^{-\tau H}\right| \phi\right\rangle=0 \quad \text { for any } \tau .
\label{no contribution}
\end{equation}

Paths that result from such a walker have equal probability of being in either half of the Slater-determinant space. Computed analytically, they would cancel, but without any knowledge of $\mathcal{N}$, they continue to be sampled in the random walk and become Monte Carlo noise.

The decay of the signal-to-noise ratio, i.e., the decay of the average sign of $\left\langle\psi_T \mid \phi\right\rangle$, occurs at an exponential rate with imaginary time.
To eliminate the decay of the signal-to-noise ratio, we impose the constrained path approximation. It requires that each random walker at each step has a positive overlap with the trial wave function $\left|\psi_T\right\rangle$ :

\begin{equation}
\left\langle\psi_T \mid \phi_k^{(n)}\right\rangle>0 .
\label{constrained path approximation}
\end{equation}

This yields an approximate solution to the ground-state wave function, $\left|\psi_0^c\right\rangle=\Sigma_\phi|\phi\rangle$, in which all Slater determinants $|\phi\rangle$ satisfy Eq.~\ref{no contribution}. From Eq.~\ref{constrained path approximation}, it follows that this approximation becomes exact for an exact trial wave function $\left|\psi_T\right\rangle=\left|\psi_0\right\rangle $. 

The details of hyper-parameters in CPQMC calculations are given as follows. The number of walkers is 1000, the number of blocks for relaxation is 10, the number of blocks for growth estimate is 3, the number of blocks after relaxation is 10, the number of steps in each block is 640, the Trotter step size is 0.01, the growth-control energy estimate is -50. To ensure the stability of the data, we intentionally increased the number of steps in each block is 640, and reduced the Trotter step size is 0.01.

% \begin{thebibliography}{}
 
% \bibitem{ALF2} ALF Collaboration, et al., arXiv:2012.11914 (2021).\
% \end{thebibliography}

% \bibliographystyle{apsrev4-1.bst}
\bibliography{ref}